\documentclass[lettersize,journal]{IEEEtran}
\usepackage{amsmath,amsfonts}
\usepackage{algorithmic}
\usepackage{algorithm}
\usepackage{array}
\usepackage[caption=false,font=normalsize,labelfont=sf,textfont=sf]{subfig}
\usepackage{textcomp}
\usepackage{stfloats}
\usepackage{url}
\usepackage{verbatim}
\usepackage{graphicx}
\usepackage{cite}
\usepackage{multirow}
\usepackage{makecell}
\usepackage{txfonts}
\usepackage{threeparttable}
\usepackage{float}

\usepackage{xcolor}
\usepackage[normalem]{ulem}

\newcommand{\add}[1]{#1}
\newcommand{\rem}[1]{}

\begin{document}

\title{Deep Learning Waveform Channel Modeling for Wideband Optical Fiber Transmission: Model Comparisons, Challenges and Potential Solutions}

\author{Minghui Shi, Hang Yang, Zekun Niu, Chuyan Zeng, Yunfan Zhang, Junzhe Xiao, Mingzhe Chen, Weisheng Hu, and Lilin Yi
\thanks{This paragraph of the first footnote will contain the date on which you submitted your paper for review, which is populated by IEEE. This work was supported by the National Key R\&D Program of China (2023YFB2905400), National Natural Science Foundation of China (62025503), and Shanghai Jiao Tong University 2030 Initiative. (Minghui Shi and Hang Yang contributed equally to this work). (Corresponding author: Lilin Yi and Zekun Niu)}
\thanks{The authors are with the State Key Laboratory of Photonics and Communications, School of Integrated Circuits (School of Information Science and Electronic Engineering), Shanghai Jiao Tong University, Shanghai 200240, China (email: minghuishi@sjtu.edu.cn; hangyang@sjtu.edu.cn; zekunniu@sjtu.edu.cn; z-mary@sjtu.edu.cn; yunfanzhang@sjtu.edu.cn; xiaojunzhe@sjtu.edu.cn; cmz1461404680@sjtu.edu.cn; wshu@sjtu.edu.cn; lilinyi@sjtu.edu.cn).}

}



\maketitle

\begin{abstract}
\add{Fast and accurate waveform simulation is critical for characterizing optical fiber channel behavior, developing digital signal processing (DSP) algorithms, optimizing optical network configurations, and advancing optical transmission systems towards wideband. Deep learning (DL) has emerged as a promising approach for waveform modeling, offering accuracy comparable to the traditional split-step Fourier method (SSFM) with significantly lower computational complexity. Despite substantial progress, most existing DL schemes are designed for few-channel, low-rate wavelength-division multiplexing (WDM) systems. Moreover, a lack of standardized accuracy evaluation methods hinders fair comparison among DL schemes, slowing their development for wideband applications.
In this paper, we propose a DSP-assisted accuracy evaluation method integrated with nonlinear DSP, which provides a fair benchmark for assessing modeling accuracy. Using this method, we conduct a comprehensive comparison of DL schemes, spanning from simple few-channel, low-rate configurations to complex wideband scenarios. The results demonstrate that FDD-based schemes combining with temporal neural networks achieve superior modeling accuracy, especially in wideband systems. Despite their advantages, we observe that even FDD-based models experience accuracy degradation as the number of channels and symbol rate increase. We analyze the underlying challenges from the perspectives of higher sampling rates, enhanced dispersion-induced inter-symbol interference, and more complex nonlinear interactions. Potential solutions are discussed, including the integration of additional physical priors and structural optimization of DL models. We believe that the standardized evaluation framework, comparative results, and in-depth challenge analysis presented in this work will facilitate the rapid advancement of DL-based channel modeling in wideband systems and support the evolution of next-generation optical networks.}

\end{abstract}


\section{Introduction} 
\IEEEPARstart{T}{he}  growing demand for data traffic is driving the evolution of optical fiber communication towards wideband, high-rate, and large-capacity systems, expanding the spectral coverage beyond the traditional C-band \cite{Multi-Band1, Multi-Band2, Multi-Band3, Multi-Band4, Multi-Band5}. Accurate, reliable, and fast optical fiber transmission simulation systems are crucial for optimizing optical networks \cite{network1, network2, network3, network4, network5, network6}, advancing digital signal processing (DSP) algorithms \cite{nonlinear1, nonlinear2, nonlinear3, nonlinear4, nonlinear5, LDSP}, and enabling end-to-end (E2E) optimization \cite{e2e1, e2e2, e2e3, e2e4, e2e5, e2e6, e2e7}. Fiber channel modeling plays a pivotal role in these simulation systems, providing critical insights into the signal evolution process within optical fibers. The propagation of signals through optical fibers is governed by the nonlinear Schrödinger equation (NLSE) \cite{Book_OFC}. Except for a few special cases, the NLSE lacks an analytical solution and must be solved through numerical simulation.

Gaussian noise (GN) \cite{GN_Model}and enhanced Gaussian noise (EGN) models \cite{EGN}offer accurate and fast fiber channel modeling techniques, primarily focusing on power-level simulations. These models treat fiber nonlinearities as Gaussian noise and estimate the generalized signal-to-noise ratio (GSNR). However, GN-like models fail to provide detailed signal waveform information, limiting their applicability in DSP algorithm design and optimization, such as nonlinearity compensation. The split-step Fourier method (SSFM) \cite{Book_NFO} is a waveform-level modeling technique for optical fiber channels. It divides the entire optical fiber link into multiple small segments, addressing linear and nonlinear effects separately within each segment. However, SSFM requires numerous iterations, leading to high computational complexity that typically scales with the fourth power of bandwidth \cite{SSFM_four_power_growth}. For wideband, long-haul systems, this computational inefficiency becomes especially pronounced, resulting in simulation times of several hours for a single parameter. Therefore, an efficient waveform-level channel modeling tool is critical to support the evolving demands of wideband optical communication systems.

Recently, deep learning (DL) techniques have gained significant attention for optical fiber channel waveform modeling, offering a promising balance between accuracy and complexity. This is attributed to their superior nonlinear fitting capabilities \cite{universal_approximators} and efficiency in parallel computation \cite{parallel}. A bidirectional long short-term memory (BiLSTM) \cite{BiLSTM} has been proposed to model 10 to 80 km optical fiber channel in intensity modulation and direct detection (IMDD) systems. A conditional generative adversarial network (CGAN) \cite{CGAN} has been proposed in a coherent systems with 1000 km transmission. The multi-head attention mechanism \cite{multi-attention, Zang:22}, Fourier neural operators (FNO) \cite{FNO_1} and center-oriented LSTM (Co-LSTM) \cite{Co_lstm} have been proposed to model long-haul transmission in a distributed manner by cascading multiple DL models in single-channel systems. A physics-informed neural operator (PINO) \cite{PINO} has been introduced to reduce the requirements of training dataset by integrating physical principles. A fully connected neural network (FCNN) \cite{DNN} and FNO \cite{FNO_5} have been introduced to achieve accurate channel modeling in few-channel WDM systems. A feature decoupled distributed (FDD) scheme \cite{FDD, ACP} combining DL models and physical models to handle linear and nonlinear effects respectively, has been extended to multi-channel WDM systems. The deep operator network (DeepONet) \cite{DPNO} scheme has been applied to wideband WDM configurations covering the C+L band, although it neglects inter-channel nonlinear effects. \add{A sequence to sequence-based FDD (Seq2Seq-FDD) scheme} \cite{Shi:25} \add{has been extended to modeling at 140 GBaud.} \cite{OSNR_impact, embedding, fiber_type} have investigated the influence of different system parameters on the accuracy and generalization ability of DL models. Beyond optical fiber transmission systems, DL technologies have also been applied to other systems, such as multi-core and few-mode optical fiber link \cite{Multi_core, few_mode}, radio over fiber (RoF) link \cite{RoF}, orthogonal frequency division multiplex (OFDM) systems \cite{OFDM}, free space optical communications (FSO) \cite{FSO} and nonlinear photonics \cite{NN1, NN2, NN3, NN4, NN5, NN6, NN7, NN8, NN9, NN10, NN11}. 

\add{DL methods present a viable alternative to traditional waveform-level channel modeling techniques. However, these approaches are typically designed for few-channel and low-rate scenarios, where the number of WDM channels is fewer than 5, and the transmission rate is below 100 GBaud. Thus they face challenges in meeting the requirements of high-speed and multi-channel optical systems. These wideband configurations represent not only the development trend of next-generation optical communications but also a regime where the traditional SSFM encounters severe computational limitations} \cite{SSFM_four_power_growth}. \add{Although DL-based schemes have potential to extend to wideband systems, a lack of standardized accuracy evaluation methods hinders fair comparison among DL schemes, slowing their development for wideband applications.}


\begin{table}[!t]
\centering
\caption{Main acronym}
\fontsize{9}{11}\selectfont  
\begin{tabular}{ll}
\hline\hline
{Acronym} & {Definition} \\
\hline
\add{ASE} & \add{amplified spontaneous emission} \\
\add{BiLSTM} & \add{bidirectional long short-term memory} \\
\add{CD} & \add{chromatic dispersion} \\
\add{CDC} & \add{chromatic dispersion compensation} \\
\add{CGAN} & \add{conditional generative adversarial network} \\
\add{CNLSE} & \add{coupled nonlinear Schrödinger equation} \\
\add{Co-LSTM} & \add{center-oriented long short-term memory} \\
\add{CPR} & \add{carrier phase recovery} \\
\add{CUT} & \add{channel under test} \\
\add{DBP} & \add{digital backpropagation} \\
\add{DeepONet} & \add{deep operator network} \\
\add{DL} & \add{deep learning} \\
\add{DP-16QAM} & \add{dual-polarized 16 quadrature amplitude modulation} \\
\add{DSP} & \add{digital signal processing} \\
\add{E2E} & \add{end-to-end} \\
\add{EDFA} & \add{erbium-doped fiber amplifier} \\
\add{EGN} & \add{enhanced gaussian noise} \\
\add{ESNR} & \add{Effective signal-to-noise ratio} \\
\add{FCNN} & \add{fully connected neural network} \\
\add{FDD} & \add{feature decoupled distributed} \\
\add{FNO} & \add{Fourier neural operators} \\
\add{FWM} & \add{four-wave mixing} \\
\add{GN} & \add{gaussian noise model} \\
\add{GSNR} & \add{generalized signal-to-noise ratio} \\
\add{IMDD} & \add{intensity modulation direct detection} \\
\add{ISI} & \add{inter-symbol interference} \\
\add{LR} & \add{learning rate} \\
\add{NLSE} & \add{nonlinear Schrödinger equation} \\
\add{NMSE} & \add{normalized mean square error} \\
\add{PINO} & \add{physics-informed neural operator} \\
\add{RNN} & \add{recurrent neural network} \\
\add{RRC} & \add{root-raised cosine} \\
\add{Rx} & \add{receiver} \\
\add{Seq2Seq} & \add{sequence to sequence} \\
\add{SL1} & \add{smooth L1} \\
\add{SPM} & \add{self-phase modulation} \\
\add{SPS} & \add{samples per symbol} \\
\add{SSFM} & \add{split-step Fourier method} \\
\add{Tx} & \add{transmitter} \\
\add{WDM} & \add{wavelength division multiplexing} \\
\add{XPM} & \add{cross-phase modulation} \\
\hline\hline
\end{tabular}
\end{table}

\add{To accelerate progress in this field and more effectively address accuracy limitations under wideband conditions, we propose a DSP-assisted accuracy evaluation method that provides a fair and comprehensive assessment of DL models. This method evaluates modeling accuracy using multiple metrics, including waveform modeling errors and transmission performance errors, with the SSFM as the reference.
Waveform modeling error is quantified using the normalized mean square error (NMSE). For transmission performance evaluation, we introduce a nonlinear compensation algorithm to validate the model’s accuracy in capturing fiber nonlinear characteristics. This approach overcomes the inconsistency between NMSE and transmission performance results that arise when only linear compensation is applied. By aligning both metrics, we achieve their unification and establish a consistent and fair evaluation baseline.
Using this evaluation framework, we design a series of increasingly complex scenarios to comprehensively compare the performance of various DL schemes, including: overall and distributed schemes, pure data-driven and data-physic hybrid-driven schemes, as well as neural networks and neural operators. 
The comparison results show that the scheme combining the FDD framework with a temporal neural network achieves an NMSE improvement of 85.6\% and 78.2\% over FNO and DeepONet, respectively, in the 5-channel, 50 GBaud scenario. In more challenging configurations—13-channel at 50 GBaud and 5-channel at 100 GBaud—FDD-BiLSTM outperforms non-FDD-BiLSTM by 83.7\% and 59.3\% in NMSE, respectively. Among all evaluated schemes, the enhanced FDD variant, Seq2Seq-FDD, achieves the best performance, exhibiting Q-factor error improvements of 0.64 dB and 0.12 dB over FDD-BiLSTM under the two wideband configurations.
Despite these advantages, we observe that the NMSE of Seq2Seq-FDD increases by  3.9 times when the number of channels rises from 5 to 25, and by 3.8 times when the symbol rate increases from 50 GBaud to 200 GBaud. This indicates that current DL models, including state-of-the-art variants, still face challenges in maintaining high accuracy in wideband scenarios.
We analyze the underlying challenges from the perspectives of higher sampling rates, enhanced dispersion-induced inter-symbol interference (ISI), and more complex nonlinear interactions. Potential solutions are discussed, including the integration of additional physical priors and structural optimization of DL models.
We believe that the standardized evaluation framework, comprehensive comparative results, and in-depth challenge analysis presented in this work will facilitate the rapid advancement of DL-based channel modeling in wideband systems and support the evolution of next-generation optical networks.}

The rest of the paper is organized as follows. Section II describes the SSFM-based optical fiber transmission simulation systems.  Section III details the architecture of DL schemes and their training processing. Section IV discusses the limitations of current accuracy metrics and introduces the DSP-assisted accuracy evaluation method. Section V presents a comprehensive comparison of DL models across scenarios of increasing complexity and analysis the challenges in applying DL models to wideband configurations. Section VI discusses potential solutions to improve the accuracy of DL models. Finally, Section VI concludes the paper.

\add{A complete list of the acronyms used in this paper is provided in the Nomenclature.}

\section{SSFM-based optical fiber transmission simulation system}

\begin{figure*}[!t]
\centering
\includegraphics[width=7in]{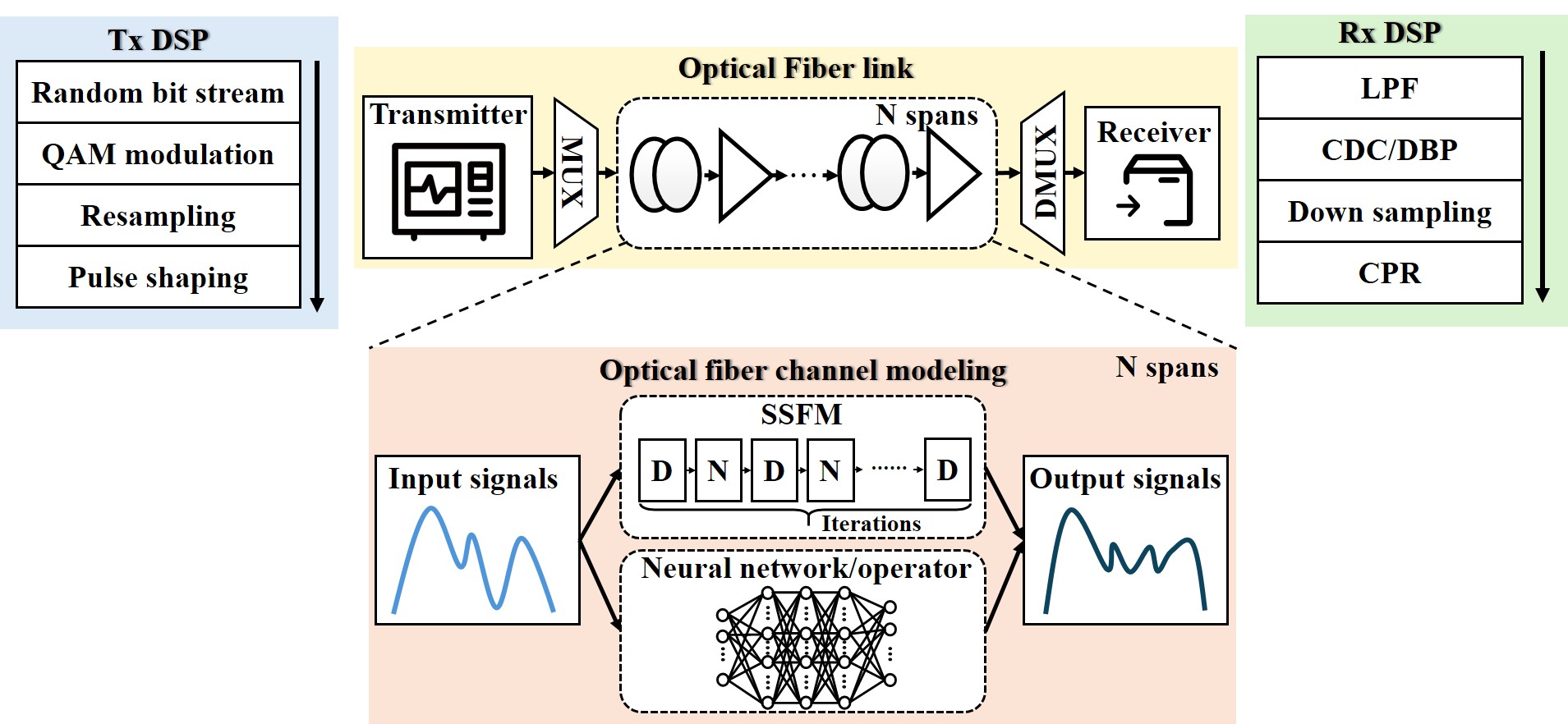}
\caption{The structure of a coherent optical transmission simulation system.}
\label{fig1}
\end{figure*}

This part introduces the structure of the SSFM-based optical fiber transmission simulation system, which generates the dataset for subsequent DL model training and serves as a baseline for accuracy evaluation.

The typical SSFM-based optical fiber transmission simulation system is shown in Fig. \ref{fig1}, including the transmitter \add{(Tx)}, optical fiber channel, and receiver \add{(Rx)}. \add{The parameters of simulation systems utilized in this paper are summarized in Table \ref{table1}.} At the \add{Tx} side, bit sequences are generated using a pseudo-random number seed and mapped to symbol sequences via dual-polarized 16 quadrature amplitude modulation (DP-16QAM). The symbol sequences are then upsampled by a factor of 4. A root-raised cosine (RRC) filter is applied for pulse shaping, with a roll-off factor of 0.1. Subsequently, the signals from different channels are modulated onto separate frequency carriers to create full-field WDM signals for transmission within the optical fiber channel. The WDM signals can be expressed as follows:
\begin{equation}
\label{euq1}
\boldsymbol{A}(z, t)=\sum_{k=1}^C \boldsymbol{A}_{\boldsymbol{k}}(z, t) \exp \left(j \Delta \omega_k t\right),
\end{equation}
where $\boldsymbol{A}$ represents the optical signals over two arbitrary orthogonal polarization modes $\boldsymbol{A_x}$ and $\boldsymbol{A_y}$. $\boldsymbol{A_k}$ is the optical signal of $k$-th channel. $z$ indicates the transmission distance and $t$ represents the time coordinate. $C$ is the total number of WDM channels, and $\Delta \omega_k=\omega_k-\omega_0$ is the difference between the central frequency of $k$-th channel and the central frequency of WDM signals. 

\begin{table}[!t]
    \centering
    \caption{Parameters of Simulation System}
    \renewcommand\arraystretch{1.5}
    \label{table1}
    \begin{tabular}{cc}
\hline\hline
{Parameters} & {Value} \\
\hline
Modulation format & DP-16QAM \\
Roll-off factor of RRC & 0.1 \\
Carrier wavelength & 1550 nm \\
Attenuation  & 0.2 dB/km \\
Dispersion  & 17 ps/(nm$\cdot$ km) \\
Nonlinear coefficient  & 1.3/(W$\cdot$ km) \\
Span length & 80 km \\
EDFA noise figure & 5 dB \\
Maximum nonlinear phase rotation of SSFM & 0.005 \\
\hline\hline
    \end{tabular}
\end{table}

The propagation of optical signal through the single-mode fiber \rem{(SMF)} is governed by the NLSE \cite{Book_NFO}, which is expressed as:
\begin{equation}
\label{euq2}
\frac{\partial \boldsymbol{A}(z, t)}{\partial z}=(\hat{\boldsymbol{D}}+\hat{\boldsymbol{N}}) \boldsymbol{A},
\end{equation}
where $\boldsymbol{D}$ is the linear operator that accounts for the effects of attenuation and chromatic dispersion (CD), and $\boldsymbol{N}$ is the nonlinear operator that \add{that represents the Kerr nonlinearity related to the signal energy.} When dealing with dual-polarization signals, the NLSE is typically considered in its coupled form, known as the Coupled NLSE (CNLSE) \cite{CNLSE}, or CNLSE simplified version, the Manakov equation \cite{Manakov1, Manakov2}. Here, we employ the Manakov equation to model the fiber channel, which can be expressed as:
\begin{equation}
i \frac{\partial \boldsymbol{A}}{\partial z}-\frac{1}{2} \beta_2 \frac{\partial^2 \boldsymbol{A}}{\partial t^2}+\frac{8}{9} \gamma|\boldsymbol{A}|^2 \boldsymbol{A}+\frac{\alpha}{2} i \boldsymbol{A}=0,
\end{equation}
where $\beta_2$ is the group velocity dispersion parameter, $\gamma$ is the nonlinear parameter and $\alpha$ is the loss parameter. \add{When dispersion and attenuation are ignored and only the nonlinear effects are considered, the NLSE can be simplified as:}

\begin{equation}
\label{nonlinear equ}
\begin{split}
\frac{\partial E_k}{\partial z} = & -j \gamma_k \left(\left|E_k\right|^2 E_k + 2 \sum_{m \neq k} \left|E_m\right|^2 E_k \right. \\
& \left. + \sum_{n, l, m \in \mathcal{H}} E_n(z, t) E_l(z, t) E_m^*(z, t) e^{j \Delta \beta_{j \lim } z} \right),
\end{split}
\end{equation}
\add{where the first term on the right-hand side represents intra-channel nonlinearity, known as self-phase modulation (SPM), while the second and third terms correspond to inter-channel nonlinearities, including cross-phase modulation (XPM) and four-wave mixing (FWM).}

The SSFM is the most commonly used numerical method for solving the NLSE. The SSFM divides long-haul optical fiber into numerous small steps, allowing linear and nonlinear operators to be considered independently. The symmetric SSFM operation \cite{symmetric_SSFM} at each step is expressed as:
\begin{equation}
\begin{aligned}
\boldsymbol{A}(z+h, t) \approx & \exp \left(\frac{h}{2} \hat{\boldsymbol{D}}\right) \exp \left\{h \hat{\boldsymbol{N}}\left[\boldsymbol{A}\left(z+\frac{h}{2}, t\right)\right]\right\} \\
& \times \exp \left(\frac{h}{2} \hat{\boldsymbol{D}}\right)\boldsymbol{A}(z, t),
\end{aligned}
\end{equation}
where $h$ is the length in each step. \add{The step size of the SSFM is determined using a maximum nonlinear phase rotation method} \cite{ssfm_nonlinear_phase} \add{to balance accuracy and complexity, which is referred to as NP-SSFM. The maximum nonlinear phase rotation is set to 0.005.} The sampling rate in SSFM is 4 times the channel number. According to their characteristics and computational convenience, the linear and nonlinear operators are calculated in the frequency and time domains, respectively. At the end of each fiber span, Erbium-doped fiber amplifiers (EDFA) are used to compensate for signal attenuation and introduce amplified spontaneous emission (ASE) noise, which can be approximated as Gaussian noise.

After transmission through the optical fiber channel, demultiplexing is performed to extract the signal in the channel under test (CUT). \rem{Receiver-side DSP}Rx DSP is then employed to compensate for signal impairments. A matched RRC filter is first applied, followed by down-sampling. Chromatic dispersion compensation (CDC) is then performed to address linear impairments. Alternatively, the CDC can be replaced by the digital backpropagation (DBP) algorithm \cite{DBP_origin}, which compensates for both linear and nonlinear impairments, providing a more thorough analysis of the nonlinear modeling capacity of DL models. Next, carrier phase recovery (CPR) and demodulation are executed. Finally, the \rem{GSNR} \add{effective signal-to-noise ratio (ESNR)} and Q-factor are calculated to evaluate the transmission performance. 

The waveforms of full-field WDM signals generated from the SSFM-based simulation system provide a rich dataset for training DL models. During testing, the same Tx and Rx process and parameters are employed to evaluate performance differences between SSFM-based and DL-based optical fiber channels.

\section{DL optical fiber channel waveform modeling}

The primary computational burden in optical fiber transmission simulation arises from the SSFM. As a result, the DL schemes primarily focus on modeling the optical fiber channel. In the following, we outline the classification of DL-based optical fiber channel modeling schemes and introduce the dataset construction and training process for the DL models.

\begin{table*}[!t]
    \centering
    \caption{DL-base waveform-level channel modeling schemes in optical fiber transmission systems}
    \renewcommand\arraystretch{1.5}
    \begin{threeparttable}
    \label{DL schemes review}
    \begin{tabular}{cccccccccc}
    \hline\hline
        Schemes & \makecell[c]{Transmission\\ mode} & \makecell[c]{Driving \\mode} & \makecell[c]{Network\\ structure} & \makecell[c]{Channel \\number} & \makecell[c]{Symbol rate\\(GBaud)} & \makecell[c]{Launch power \\(dBm/channel)} & \makecell[c]{Modulation \\format} & \makecell[c]{Distance\\(km)} & Reference \\ \hline
        BiLSTM & $\triangle$ & $\square$ & $\lozenge$ & 1 & 10 & 0$\sim$15 & PAM4 & 80 & \cite{BiLSTM} \\ 
        CGAN & $\triangle$ & $\square$ & $\lozenge$ & 1 & 30 & -2$\sim$10 & 16QAM & 1000 & \cite{CGAN} \\ 
        Multi-head Attention & $\blacktriangle$ & $\square$ & $\lozenge$ & 1 & 40 & 3&16QAM & 1000 & \cite{multi-attention, Zang:22} \\ 
        FNO & $\blacktriangle$ & $\square$& $\blacklozenge$ & 1 & 28 & -5$\sim$5 &64QAM & 1200 & \cite{FNO_1} \\ 
        PINO & $\blacktriangle$ & $\blacksquare$ & $\blacklozenge$ & 1 & 14 & -3$\sim$3&16QAM & 320 & \cite{PINO} \\
        FCNN & $\blacktriangle$ & $\square$ & $\lozenge$ & 2 & 30 & 4&16QAM & 80 & \cite{DNN} \\ 
        FNO & $\blacktriangle$ & $\square$ & $\blacklozenge$ & 5 & 30 & -2$\sim$2&DP-16QAM & 800 & \cite{FNO_5} \\ 
        FDD-BiLSTM & $\blacktriangle$ & $\blacksquare$ & $\lozenge$ & 41 & 30 & -1&DP-128QAM & 1040 & \cite{FDD} \\ 
        FDD-Co-LSTM & $\blacktriangle$ & $\blacksquare$ & $\lozenge$ & 1 & 32 & 0&DP-16QAM & 2800 & \cite{Co_lstm} \\ 
        FDD-Transformer & $\blacktriangle$ & $\blacksquare$ & $\lozenge$ & 21 & 30 & 2.5&DP-16QAM & 1200 & \cite{ACP} \\ 
        DeepONet & $\blacktriangle$ & $\square$ & $\blacklozenge$ & 96 & 100 & 0&DP-16QAM & 800 & \cite{DPNO} \\
        Seq2Seq-FDD & $\blacktriangle$ & $\blacksquare$ & $\lozenge$ & 5 & 140 & 3.5$\sim$8.5&DP-16QAM & 1200 & \cite{Shi:25} \\
        \hline\hline
    \end{tabular}
    \begin{tablenotes}    
        \item[1] Overall: $\triangle$ \enspace Distributed: $\blacktriangle$ \qquad Pure data-driven: $\square$ \enspace Data-physic hybrid-driven:  $\blacksquare$ \qquad Neural network: $\lozenge$  \enspace Neural operator:  $\blacklozenge$
      \end{tablenotes}           
    \end{threeparttable}
\end{table*}

\subsection{Classification of DL schemes}

DL schemes utilize parameterized neural networks or operators to uncover hidden correlations and patterns from datasets generated by SSFM-based simulations, eliminating the need for complex prior mathematical and physical knowledge. \rem{Advanced DL schemes have achieved efficient optical fiber channel waveform modeling.} In this section, we provide a comprehensive review of DL schemes. To offer a clear and detailed introduction and comparison of various DL schemes, we have categorized them into three classes based on their distinct characteristics, including transmission modes, driving modes, and network architectures.

\begin{itemize}
\item Transmission mode refers to whether a single model or multiple models are used to model long-distance optical fiber link transmission. This can be divided into overall and distributed schemes.
\item Driving mode refers to whether physical knowledge is integrated alongside data-driven approaches. This can be classified into pure data-driven or physics-data hybrid-driven schemes.
\item Network structure refers to the differences between various parameterized architectures, which can be divided into neural networks and neural operators.
\end{itemize}
\subsubsection{\textbf{Overall schemes and distributed schemes}}

\add{Long-haul optical transmission links are composed of a cascade of multiple fiber spans and optical amplifiers. To model such links, DL schemes generally adopt two main approaches: overall schemes and distributed schemes, which refer to the use of a single DL model or multiple DL models, respectively, to represent the entire transmission link. }

Overall schemes treat the fiber link as a single integrated module and employ a single network to model the entire transmission link. Examples of this approach include the overall BiLSTM \cite{BiLSTM} and the overall CGAN \cite{CGAN}. The overall BiLSTM has demonstrated high accuracy in IMDD systems over transmission distances ranging from 10 to 80 km. However, it is limited to single-span scenarios due to the deterministic nature of the BiLSTM, which struggles to effectively capture the random noise introduced by EDFA. To enable long-haul transmission modeling over 1000 km, a generative model, the CGAN, has been proposed. CGAN can learn and reproduce specific data distributions, thereby addressing the challenge of modeling stochastic impairments. However, the adversarial training process in CGAN is prone to instability and may fail to converge to optimal performance \cite{GAN_challenge, gan_review}, which particularly limits its applicability in multi-channel WDM systems.

\add{To enhance the long-haul modeling capability, distributed schemes were proposed. In these schemes, the entire optical fiber link is decomposed into multiple segments rather than treated as a single integrated module, with each DL model responsible for modeling one fiber span. This architecture enables more flexible handling of stochastic impairments, such as ASE noise, across spans. Related studies adopting this approach include FNO, DeepONet, and FDD-BiLSTM, among others.  In these works, each model uses either identical or slightly fine-tuned parameters to capture the response of a single fiber span, and multiple models are cascaded to achieve long-haul transmission.} Distributed schemes have demonstrated robust performance in both single- and multi-channel WDM systems. By focusing on the characteristics of a single fiber span, these schemes avoid the complexities associated with modeling the entire fiber link, such as the accumulation of linear and nonlinear effects over distance and the amplifier noise, which is challenging for deterministic DL models. This simplification of features enables faster convergence and improved accuracy in DL models. Additionally, distributed schemes allow for flexible adjustments in transmission length by varying the number of cascaded models. However, the iterative cascading of multiple models introduces inevitable iterative errors, which can degrade performance in long-haul transmission. Some studies address these iterative errors by constructing multi-span datasets for training DL models \cite{ACP} or by fine-tuning the models used in later spans \cite{Co_lstm, PINO}. Despite challenges posed by iterative errors, distributed schemes offer higher accuracy and flexibility compared to overall schemes. As such, they are seen as potential mainstream methods for fiber channel modeling in wideband and long-haul optical transmission systems.

\subsubsection{\textbf{Pure data-driven schemes and data-physics hybrid-driven scheme}}

DL schemes directly learn the mapping between input and output signals generated from SSFM-based simulations and do not require extensive prior physical knowledge, characterizing them as pure data-driven schemes. To improve modeling accuracy, some studies integrate principles into the modeling framework, giving rise to data-physics hybrid-driven schemes.

Pure data-driven schemes rely solely on training data from SSFM simulations. During training, loss functions such as mean squared error or smooth L1 (SL1) loss are employed to optimize the model parameters, enabling convergence to the features present in the data. These schemes are valued for their structural simplicity and straightforward training pipeline. However, they face the challenge of requiring the model to learn all characteristics of the data, including complex linear, nonlinear, and random effects within the optical fiber channel. Furthermore, pure data-driven models rely solely on waveform errors as loss functions, which may lead to suboptimal solutions that fail to satisfy the underlying physical laws, such as the NLSE constraint.

To enhance the accuracy of pure data-driven methods, data-physics hybrid-driven schemes have been proposed, which incorporate physical principles into the model architecture, preprocessing, or training process. Notable examples include the FDD and PINO frameworks. The FDD scheme separates optical channel characteristics into linear and nonlinear components. Linear effects are modeled using a physical model, while a NN is responsible for capturing nonlinear effects. This design ensures accurate modeling for linear effects, reduces the complexity of the learning task of NN, and improves the accuracy of nonlinear modeling. \add{Yang et al.} \cite{FDD} \add{integrated FDD with BiLSTM, successfully extending the framework to long-haul WDM systems. Shi et al.} \cite{ACP} \add{combined FDD with the Transformer architecture, enabling accurate modeling in 21-channel WDM configurations. Zheng et al.} \cite{Co_lstm} \add{improved the BiLSTM structure within FDD by proposing a Co-LSTM, which adopts a multi-input multi-output mode during inference, leveraging the recurrent mechanism of LSTM to generate multiple output symbols per step, thereby reducing computational complexity. Building on this work, Shi et al.} \cite{Shi:25} \add{further optimized the FDD-Co-LSTM architecture and proposed Seq2Seq-FDD, which introduces transfer learning and incorporates the multi-input multi-output mode into the training phase. This alleviates parameter mismatch and enhances training efficiency and model accuracy, enabling accurate modeling in 140-GBaud systems.}
The PINO scheme integrates physical principles by embedding the NLSE into the loss function of a DeepONet-based channel model. This enables physics-informed or even unsupervised training, significantly reducing the reliance on large labeled datasets and improving training efficiency. Compared to pure data-driven methods, both FDD and PINO demonstrate superior accuracy in both single-channel and multi-channel WDM systems. Additionally, hybrid models mitigate the risk of suboptimal solutions and ensure better adherence to the physical properties described by the NLSE, enhancing the reliability of DL-based channel modeling.

\subsubsection{\textbf{Neural network schemes and neural operator schemes}}

DL schemes based on neural networks have achieved outstanding performance in optical fiber channel modeling. In recent years, deep neural operators have emerged as a novel DL paradigm, advancing rapidly and demonstrating strong generalization capabilities in solving partial differential equations  \rem{(PDEs)} \cite{operator_theory1, operator_theory2, operator_theory3}.

\add{Neural networks, grounded in the universal approximation theorem, can approximate any continuous function to arbitrary precision given sufficient depth and capacity. These networks learn mappings between finite-dimensional, discrete vector spaces. Various architectures have demonstrated high accuracy in modeling optical fiber channels, among which temporal models—particularly the BiLSTM and the Transformer—have shown superior performance.} BiLSTM, a representative recurrent neural network, captures temporal dependencies through its forward and backward hidden state propagation. The Transformer leverages a self-attention mechanism, enabling global temporal feature extraction and inherent parallelization, which enhances training efficiency and modeling capability.

Neural operators represent a paradigm shift by learning mappings between infinite-dimensional function spaces—offering a more general formulation than the discrete vector space mappings typical of conventional neural networks. Two representative neural operators, the FNO and DeepONet, have been introduced for optical fiber channel modeling. FNO employs Fourier transform techniques to construct solution operators for the NLSE from both time- and frequency-domain perspectives. DeepONet, composed of a branch network and a trunk network, excels at capturing complex functional relationships between inputs and outputs. It can incorporate the NLSE into the loss function as a physics-informed constraint, thereby improving model convergence. The input of the trunk network includes distance and time coordinates, which support automatic differentiation. Through a stacked DeepONet architecture, it has been successfully applied to fully loaded C+L-band WDM systems. The primary advantage of neural operators is their superior generalization ability. The goal of neural operators is to approximate the mapping between function spaces, and once this mapping is learned, the parameterized operator can consistently locate the correct solution in the objective function space, regardless of changes in initial conditions. This generalization ability has proven effective for varying launch power in optical fiber channel modeling.	

Overall, neural networks with efficient architectural designs and neural operators with enhanced generalization offer complementary strengths in fiber channel modeling. Future research should focus on integrating the representational power of neural networks with the functional generalization of neural operators to further advance their application in wideband, long-haul optical communication systems.

\begin{figure*}[!t]
\centering
\includegraphics[width=7in]{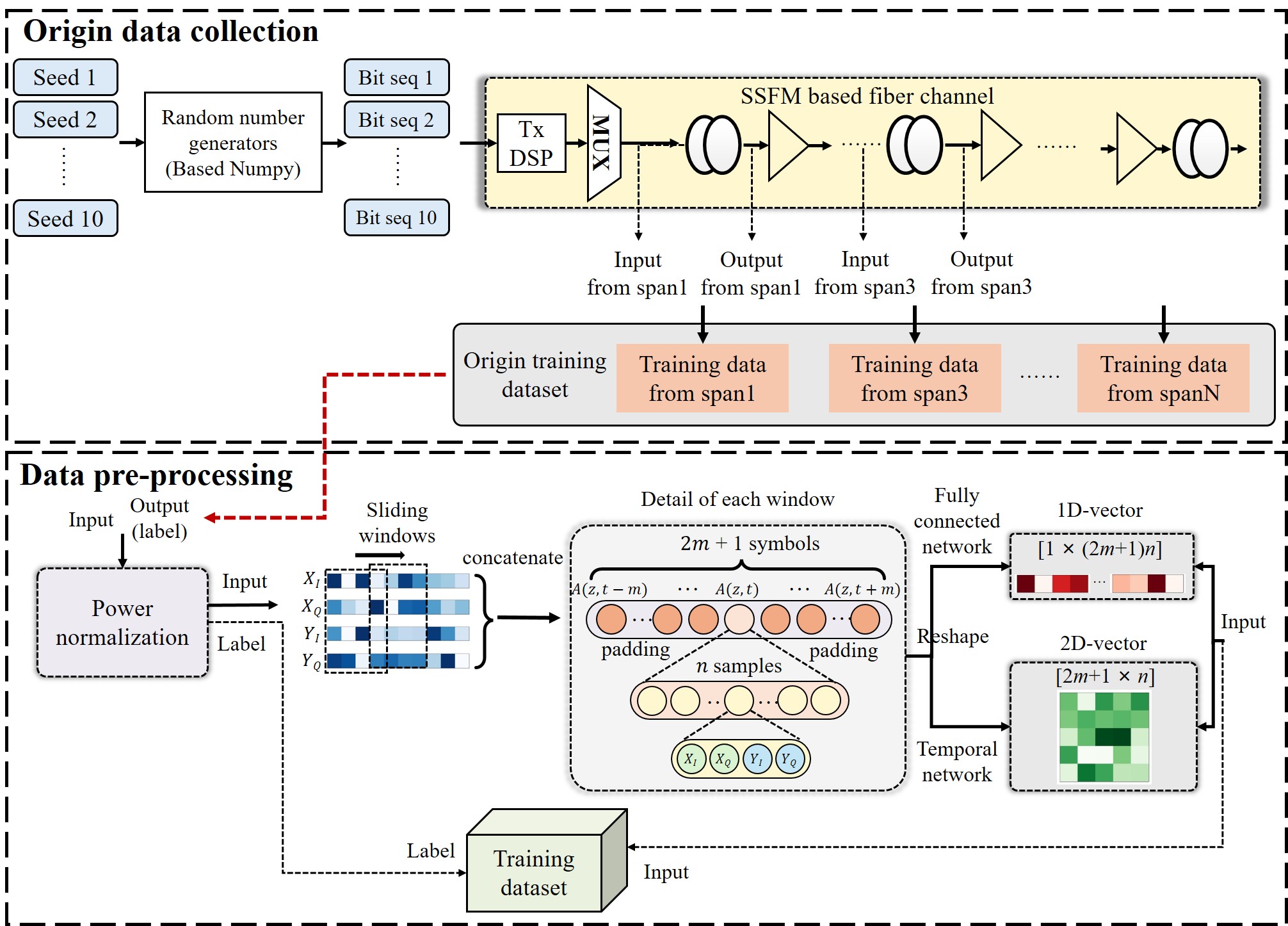}
\caption{The processing of dataset construction.}
\label{fig2}
\end{figure*}

\subsection{Dataset construction}
Before training the DL models, we need to construct the dataset, which includes two stages: original data collection and data pre-processing, as shown in Fig. \ref{fig2}. The original data is collected from SSFM-based simulation systems. To mitigate overfitting and reduce pattern prediction by neural networks \cite{overfitting1, overfitting2}, multiple random seeds are employed to generate diverse bit sequences through a NumPy-based random number generator. \rem{In this study, ten distinct random seeds are employed to generate the training dataset, and one different random seed is used to generate the testing dataset.} Each generated bit sequence is mapped to the DP-16QAM constellation, producing sequences of 7680 symbols per polarization, which are then transmitted through an SSFM-based optical fiber channel. \add{In this paper, the length $L_S$ of each fiber span is fixed at 80 km, while the number of fiber spans is determined based on the total transmission distance $D$.} The data collection methods differ between overall and distributed schemes. In the overall scheme, a single DL model is employed to model the long-haul optical fiber channel, collecting waveform data only at transmission distances of 0 km and $D$ km (the final transmission distance). On the contrary, the distributed scheme uses multiple cascaded DL models to model each fiber span individually. Thus, input and output data from various fiber spans are collected. In this study, the waveform data is collected from fiber spans [1, 3, 5, 7, 9] to enrich the training dataset, incorporating a variety of span input and output features, which enhances the model’s performance for long-haul transmission \cite{ACP}. ASE noise, caused by the EDFA, is introduced into the output signal after each fiber span transmission, contributing a random property to the input-to-output mapping. Only generative models, such as CGAN, can handle random characteristics. Other neural networks, such as FCNN or BiLSTM, are typically capable only of fitting deterministic mappings and struggle with random features. To address this issue, we collect the output waveform before the EDFA introduces random noise, ensuring that the input-output mapping in the dataset contains only deterministic features.

After collecting original waveform data, data preprocessing is necessary to accelerate model convergence and adapt to different input format requirements across various DL models. For all DL models, the preprocessing step includes power normalization, sliding windows, and reshaping. Additionally, linear feature decoupling is necessary for the FDD scheme. The output signals are passed through simple linear compensation to eliminate CD, which can be described by the following equation:
\begin{equation}
\boldsymbol{A}(z-L, \omega)=A(z, \omega) \exp \left(-i \frac{\beta_{2}}{2} \omega^{2}\right)(-L)
\end{equation}
After linear compensation, the output signals contain only nonlinear characteristics. Then, power normalization is applied to facilitate model convergence by adjusting the input and output signals according to the following equation \cite{Book_OFC}:
\begin{equation}
\begin{aligned}
\tilde{x} = x \cdot \sqrt{\frac{1}{\sum_{i=1}^{N_{\mathrm{ch}}} P_i}},
\end{aligned}
\end{equation}
where $N_{ch}$ represents the total number of the WDM channels, and $P_i$ is the power of the $i$-th channel. Given that the dataset comprises complex signals with dual polarization (X and Y) and DL models are better at dealing with real number than complex ones. The complex signals are decomposed into two real-valued components: in-phase (I) and quadrature (Q) signals. At each time step, the real-valued signals of the two polarizations—denoted as $X_I$, $X_Q$, $Y_I$ and $Y_Q$—are concatenated into a one-dimensional vector. To preserve accuracy, it is essential to account for inter-symbol interference (ISI) induced by CD when constructing the input data. The number of symbol affected by ISI is determined using the equation:
\begin{equation}
\label{ISI}
\begin{aligned}
N_{\mathrm{ISI}} = \frac{\Delta T}{\Delta t} = \frac{L \beta_2 \Delta w}{1/S} = L \beta_2 \Delta w S,
\end{aligned}
\end{equation}
where $N_\mathrm{ISI}$ represents the number of symbols affected by ISI, $\Delta T$ is the temporal width affected by ISI, $\Delta t$ is the time duration of one symbol, $L$ is the transmission distance, $\Delta\omega$ is the spectrum width, and the $S$ is the symbol rate. To account for ISI, adjacent symbols are included in the input data using a sliding window. The window length is adjusted based on the ISI strength, which varies with the symbol rate, spectrum width, and transmission distance. The input window consists of symbols at different times, represented as $[A(z, t-m), \cdots, A(z, t), \cdots, A(z, t+m)]$, where $A(z,t)$ is a one-dimensional vector aligned at distance $z$ and time $t$, with dimension $d$. Here, $m$ is the number of adjacent symbols, typically set as $(N_\mathrm{ISI}-1)/2$, ensuring accurate representation of CD effects. The FDD scheme employs a linear decoupling method, enabling the DL model to focus exclusively on fitting nonlinear characteristics. Inter-symbol correlations in residual nonlinearities are shorter than that caused by CD, due to continuous signal power attenuation during fiber propagation. The nonlinear inter-symbol correlations can be approximated using the effective nonlinear length, expressed as:

\begin{equation}
\label{Non_ISI1}
\begin{aligned}
L_{\mathrm{eff}} = \frac{1}{\alpha},
\end{aligned}
\end{equation}

\begin{equation}
\label{Non_ISI2}
\begin{aligned}
N_{\mathrm{NL}} = L_{\mathrm{eff}} \beta_2 \Delta w S,
\end{aligned}
\end{equation}
where $L_{\mathrm{eff}}$ presents the effective nonlinear length, and $N_{\mathrm{NL}}$ denotes the number of symbols affected by nonlinear inter-symbols correlations. Each symbol within input window comprises $N$ sample points, with each sample point containing the four-dimensional signals $X_I$, $X_Q$, $Y_I$, $Y_Q$. Thus, the total dimension $d$ of $A(z,t)$ is calculated as:
\begin{equation}
d=4 N_{\mathrm{SPS}}=4 * 4 * N_{c h}
\end{equation}
where $N_{\mathrm{SPS}}$ is the samples per symbol (SPS) of SSFM, which is 4 times the channel number. Following the sliding window processing, the input window is reshaped to meet the specific input requirements of various DL models, including fully connected networks and temporal networks. For fully connected networks, such as GAN, FNO,  and the branch net of DeepONet, the input is reshaped as a one-dimensional vector containing symbols from $2m+1$ distinct time steps, represented as $[b, (2m+1)*d]$, where $b$ is the batch size. For temporal networks, such as BiLSTM and multi-head attention mechanism, the input is structured as a two-dimensional vector, expressed as $[b, 2m+1, d]$, where the second dimension corresponds to different time steps, and the third dimension represents the dimension of each symbol. Additionally, the distance $z$ and time $t$ are considered as the inputs to the trunk net of DeepONet \cite{DPNO}.

\subsection{Training processing}
\add{Here, we describe the specific training configuration of the DL model. The training dataset is generated using ten distinct random number seeds, while an eleventh, independent seed is used to generate the testing dataset. Each seed produces a sequence of 7,680 symbols per polarization, which are partitioned into 7,680 input–output pairs to form the training samples. The entire dataset is split into training, validation, and test sets in an 8:2:1 ratio. The batch size for each parameter update is set to 512.} The optimal model parameters are selected based on the test set results to avoid overfitting. \add{In this work, the SL1 loss is adopted as the loss function}, defined as

\begin{equation}
Loss= \frac{1}{N_{\mathrm{data}}}\sum_{i=1}^{N_{\mathrm{data}}}\begin{cases}0.5 (\hat{y}_i-y_i)^2 & \text { if }|\hat{y}_i-y_i|<1 \\ |\hat{y}_i-y_i|-0.5 & \text { otherwise }\end{cases},
\end{equation}
where $N_{data}$ is the number of the data size, $y$ and $\hat{y}$ denote the outputs from SSFM and DL schemes, respectively. \rem{In this study, we choose SL1 as the loss function.} The number of epochs is set at 1000. The Adam optimizer \cite{Adam} is used with an initial learning rate (LR) of 5E-4, and the LR is adjusted using a cosine annealing schedule \cite{SGDR} for each epoch to improve training performance. 

\section{The accuracy evaluation method for DL-based schemes}
We have introduced the classification and the training process of DL schemes. After training, it is important to evaluate the accuracy of DL models. This section reviews existing accuracy evaluation metrics, including NMSE, which reflects waveform errors, and constellations, ESNR, and Q-factor, which reflect transmission performance errors. Further, we find and analyze their potential limitations, including the difficulty in defining an acceptable NMSE threshold and the inconsistency between NMSE and Q-factor errors. To overcome the above shortcomings, we propose a DSP-assisted accuracy evaluation method, offering a fair and comprehensive benchmark for assessing accuracy and establishing a robust foundation for subsequent comparisons between various DL models in wideband systems.

\subsection{The metrics for accuracy evaluation}
Accuracy evaluation metrics for DL-based optical fiber channel modeling typically involve waveform errors and transmission performance errors. For waveform modeling schemes, waveform errors offer an intuitive means of accuracy comparison. These errors can be analyzed by examining deviations in waveform profiles and are quantitatively measured using the NMSE, defined as:
\begin{equation}
N M S E=\frac{\sum_{i-1}^{N_{\text {data }}}\left|\hat{y}_i-y_i\right|^2}{\sum_{i-1}^{N_{\text {data }}}\left|y_i\right|^2},
\end{equation}
where $N_{data}$ is the number of the data size, $y$ and $\hat{y}$ denote the outputs from SSFM and DL schemes, respectively. When ASE noise from the EDFA is incorporated, the same pseudo random noise is applied to both SSFM and DL schemes during NMSE calculations to maintain consistency and avoid significant discrepancies in the NMSE values. Previous studies have established that an NMSE threshold below 0.02 indicates sufficiently accurate DL models \cite{BiLSTM}.

In addition to waveform errors, transmission performance errors are crucial metrics in communication systems. To evaluate transmission performance, DSP algorithms are essential for compensating channel impairments. Constellations are commonly used to analyze transmission performance, offering a visual representation of noise distribution by showing the deviation between the Rx symbols and the Tx symbols. \add{Quantitatively, the ESNR is calculated to evaluate the sparseness of the received symbols in the constellation with respect to their references Tx symbols.} ESNR is defined as:
\begin{equation}
\mathrm{ESNR} = 10 \log_{10} \left( \frac{P_s}{\mathbb{E} |rx - tx|^2} \right),
\end{equation}
where $rx$ and $tx$ are the received and transmitted symbols, respectively, $P_s$ is the signal power. Subsequently, demodulation is performed to calculate the bit error rate (BER) and Q-factor, defined as:
\begin{equation}
B E R=\frac{N_{error}}{N_{bits }}
\end{equation}
\begin{equation}
Q=20 \log _{10}\left(\sqrt{2} e r f c^{-1}(2 B E R)\right),
\end{equation}
where $N_{error}$ is the number of erroneous bits, $N_{bits}$ is the total number of bits, and $e r f c(x)$ is complementary error function. \add{The Q-factor error is calculated as the difference between the Q-factor of the SSFM and that of the DL models, expressed as:}
\begin{equation}
Q_{\mathrm{error}}=Q_{\mathrm{SSFM}} - Q_{\mathrm{DL}}.
\end{equation}
The Q-factor is derived as a logarithmic transformation of the BER, implying that identical Q-factor errors correspond to different BER errors under varying noise levels. The accumulation of nonlinear and random noise varies across different transmission distances or WDM configurations. To ensure fair comparisons across different WDM configurations, extra noise are introduced at the Rx side to keep BER and Q-factor values consistent across different configurations. This noise, applied to the Rx signals for both SSFM and DL schemes, is generated using the pseudo same random seed. In this study, the BER is maintained at 4E-2 for calculating the Q-factor, which is the 25\% soft-decision forward-error-correction \rem{(SD-FEC)} BER threshold \cite{FEC_threshold}.

\add{When calculating these accuracy evaluation metrics, we generate the test pseudo random binary sequence (PRBS) using a pseudo random number seed distinct from that used for the training set, and mapped it to DP-16QAM symbols. This symbol sequence contains $2^{18}$ symbols per polarization. In the subsequent testing process, only one PRBS sequence is used, as we observe that the performance of the DL model remains consistent across different PRBS. For a detailed analysis of the impact of different PRBS on the test results, please refer to Appendix A.}

\subsection{The potential limitations of accuracy metrics}
The metrics discussed in the previous subsection are widely employed to evaluate the accuracy of DL-based optical fiber channel waveform modeling. However, the interrelationships between these metrics have not been thoroughly examined. We find that waveform errors do not directly correlate with transmission performance metrics, and, as a result, the NMSE threshold of 0.02 may not be universally adequate, especially in highly nonlinear systems. Additionally, the consistency between waveform errors and transmission performance errors may become problematic, particularly after long-haul transmission. In the following, we investigate these potential limitations and present illustrative examples to clarify these relationships.

\subsubsection{\textbf{The acceptable NMSE threshold}}
\begin{figure*}[!t]
\centering
\includegraphics[width=7in]{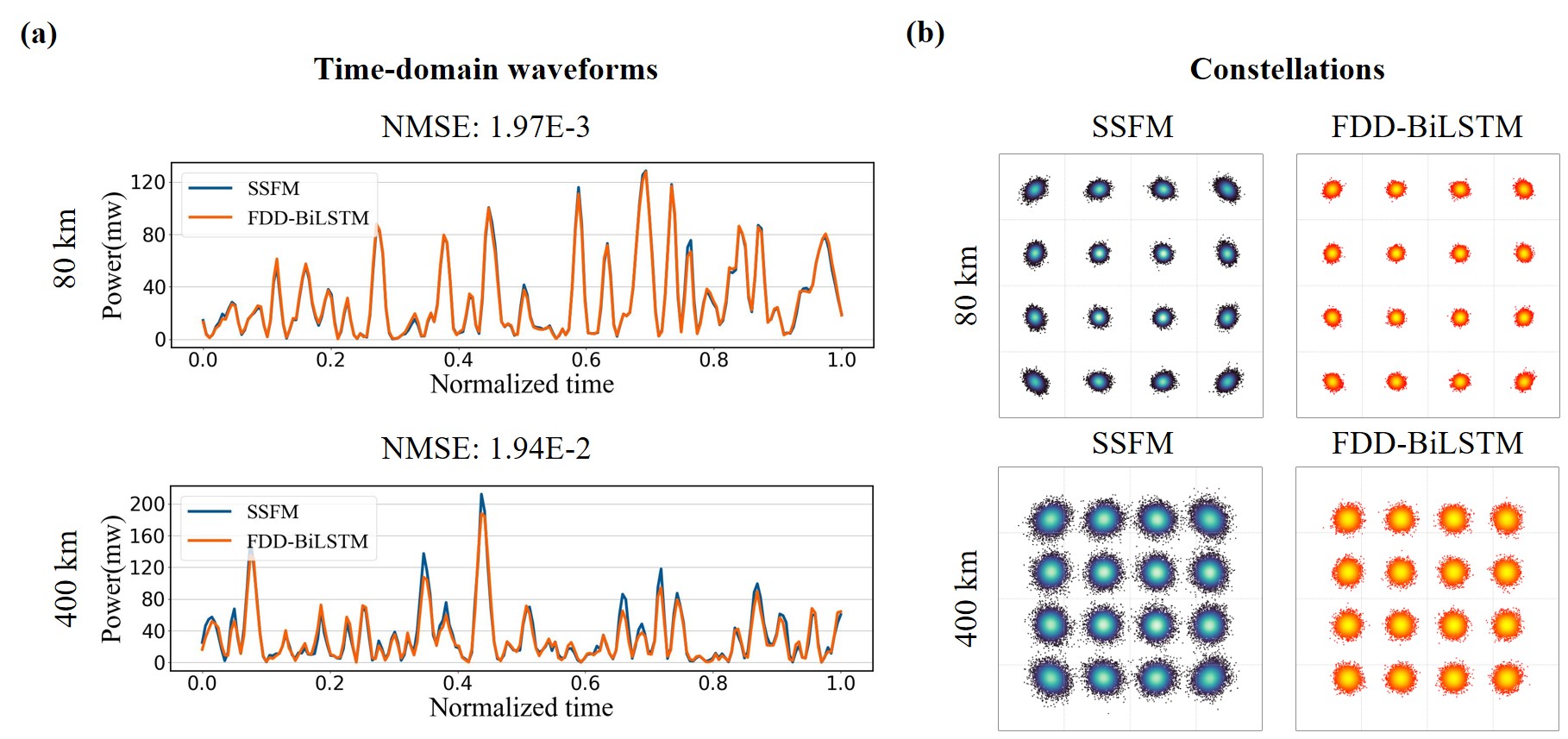}
\caption{Comparison of FDD-BiLSTM and SSFM at 80 km and 400 km. (a) Time-domain waveforms and NMSE values. (b) Constellations after linear DSP.}
\label{fig3}
\end{figure*}

\begin{figure}[!t]
\centering
\includegraphics[width=3in]{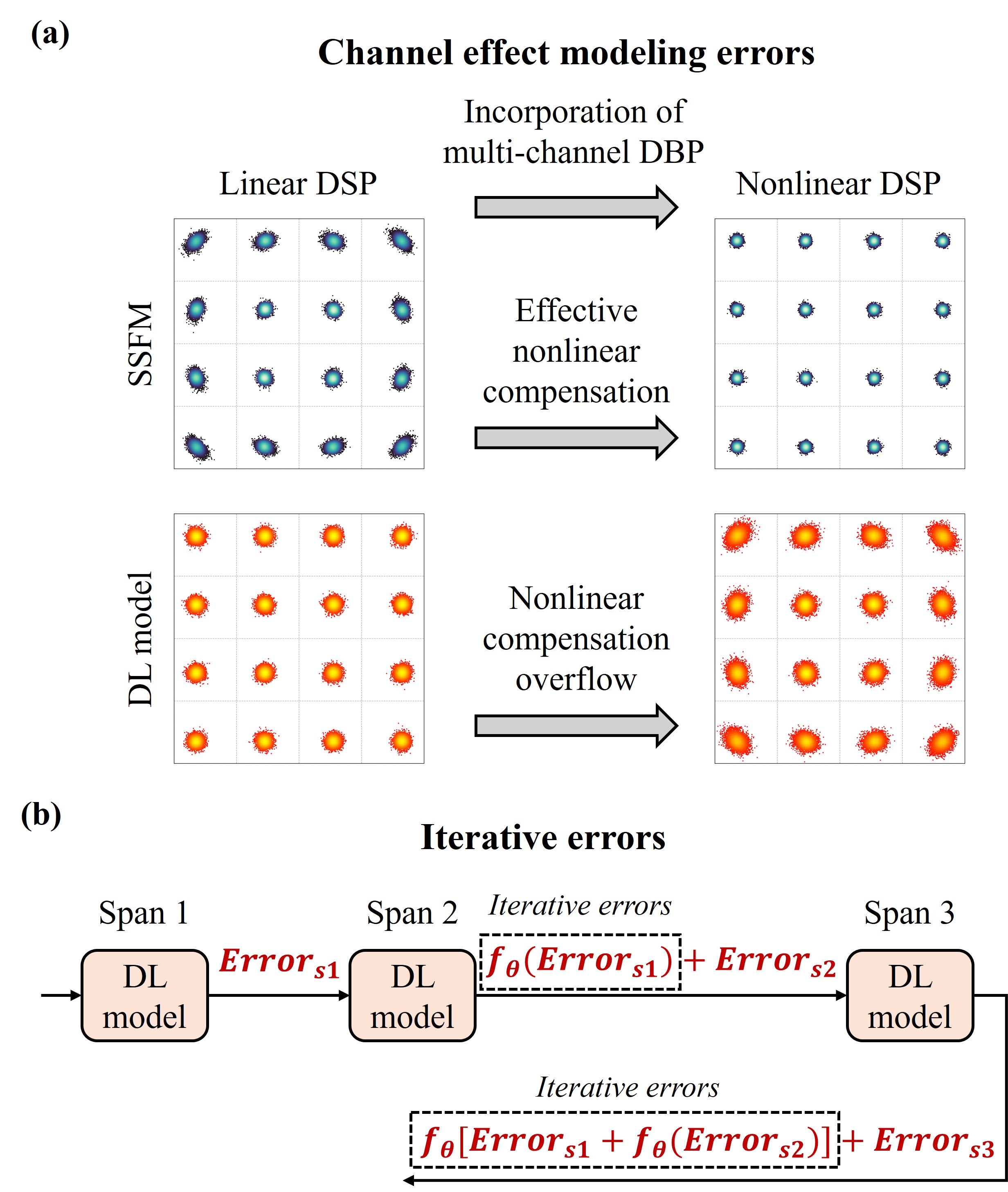}
\caption{Schematic diagrams of two types of errors. (a) Channel effect modeling errors and the influence of multi-channel DBP on them. (b) Iterative errors.} 
\label{two_errors}
\end{figure}

An NMSE threshold of 0.02 has been established as an acceptable benchmark for DL schemes in previous studies \cite{BiLSTM} and is widely adopted in subsequent research \cite{CGAN, FDD}. To evaluate the suitability of this threshold under strong nonlinear conditions, we use FDD-BiLSTM as an example and test its performance in an 11-channel configuration with a transmission rate of 140 GBaud and a launch power of 8 dBm---approximately 2.5 dB higher than the optimal launch power and situated in the high nonlinear regime. First, waveform errors are assessed by examining the time-domain waveforms and NMSE values between FDD-BiLSTM and SSFM. Fig. \ref{fig3}(a) shows the time-domain waveforms and their NMSE at both 80 km and 400 km transmission, with NMSE values of 1.97E-3 and 1.94E-2, respectively--both of which fall below the accepted threshold of 0.02. Next, transmission performance is assessed for both FDD and SSFM to explore the relationship between the acceptable NMSE threshold and transmission performance errors. Fig. \ref{fig3}(b) displays the constellations of FDD and SSFM after linear DSP at 80 km and 400 km. The constellations of FDD exhibit less noise accumulation than SSFM, particularly due to nonlinear phase rotation associated with signal power. These results highlight that even when waveform distortions remain within the acceptable NMSE threshold, discrepancies in noise distribution can be significant. The primary reason for this disparity lies in the fact that nonlinearity acts as a perturbation term \cite{perturbation}, which minimally affects waveform distortions but significantly influences noise accumulation and transmission performance. Accurate modeling of both random and nonlinear noise is essential to reflect the actual behavior of the fiber channel and support applications such as nonlinear compensation algorithm design. 

These findings indicate that the NMSE threshold of 0.02 is insufficient in certain scenarios with strong nonlinearities, and defining a universal NMSE threshold across varying scenarios remains challenging due to varying noise accumulation. Although establishing a universal NMSE threshold is difficult, NMSE continues to serve as a valuable relative metric for evaluating the accuracy of waveform-level modeling. Further accuracy evaluation should also incorporate transmission performance error to ensure that the model’s accuracy meets the specific requirements of applications, such as using DL models for fast DSP algorithm optimization or link performance prediction.

\subsubsection{\textbf{The inconsistency between waveform errors and transmission performance errors for distributed schemes}}

\add{Establishing a universal NMSE threshold is challenging, and a more comprehensive accuracy evaluation requires the incorporation of transmission performance metrics. However, transmission performance---quantified by the Q-factor or ESNR---is a statistical measure influenced by various impairments accumulated in the Rx signal after long-haul transmission. For DL models employed in distributed schemes, two distinct error sources—channel effect modeling error and iterative error—coexist and exhibit opposing effects on transmission performance. These errors tend to partially cancel each other after long-haul transmission, leading to fluctuations in transmission performance and an inconsistency between waveform error and transmission performance error, thereby obscuring the true modeling accuracy. In the following, we elaborate on the origins of these errors and their impact on transmission performance.}

\add{The left half of Fig.} \ref{two_errors}(a) \add{illustrates the channel effect modeling error between the DL model and the SSFM reference. As shown in the SSFM constellation, significant nonlinear phase rotation is observed—particularly for high-power symbols—whereas the corresponding symbols in the DL model exhibit a more Gaussian-like distribution, indicating insufficient capture of nonlinear effects. This discrepancy leads to an overestimation of the DL model’s transmission performance relative to SSFM ($Q_{SSFM}<Q_{DL}$). The second source of error is the iterative error, which arises in distributed modeling schemes due to the cascading of multiple DL models. During cascading, errors generated in earlier spans propagate and accumulate through subsequent spans. }Fig. \ref{two_errors}(b) \add{illustrates the mechanism of this error accumulation. Here, $f_{\theta}$ denotes the response function of the DL model with parameter $\theta$, and $Error_{Sk}$ represents the error introduced after signal transmission through the $k$-th span. The error from the first span, $Error_{S1}$, is transmitted through the second span, resulting in an accumulated iterative error $f_{\theta}(Error_{S1})$. This process continues into the third span, generating $f_{\theta}[f_{\theta}(Error_{S1})]$, and so on. As the transmission distance increases, iterative error accumulation becomes increasingly significant, degrading the accuracy of the DL model. Previous studies} \cite{FDD, ACP, Co_lstm} \add{have shown that such iterative error propagation is inherent in distributed frameworks and limits modeling accuracy in long-haul transmission. After long-haul transmission, both errors coexist: the channel effect modeling error leads to an overestimation of performance, while the iterative error causes underestimation. Because their impacts are opposite, they partially cancel each other, leading to fluctuations in transmission performance and an inconsistency between transmission performance and waveform errors. This masks the true modeling accuracy and hinders reliable performance evaluation.}

\begin{figure*}[!t]
\centering
\includegraphics[width=7in]{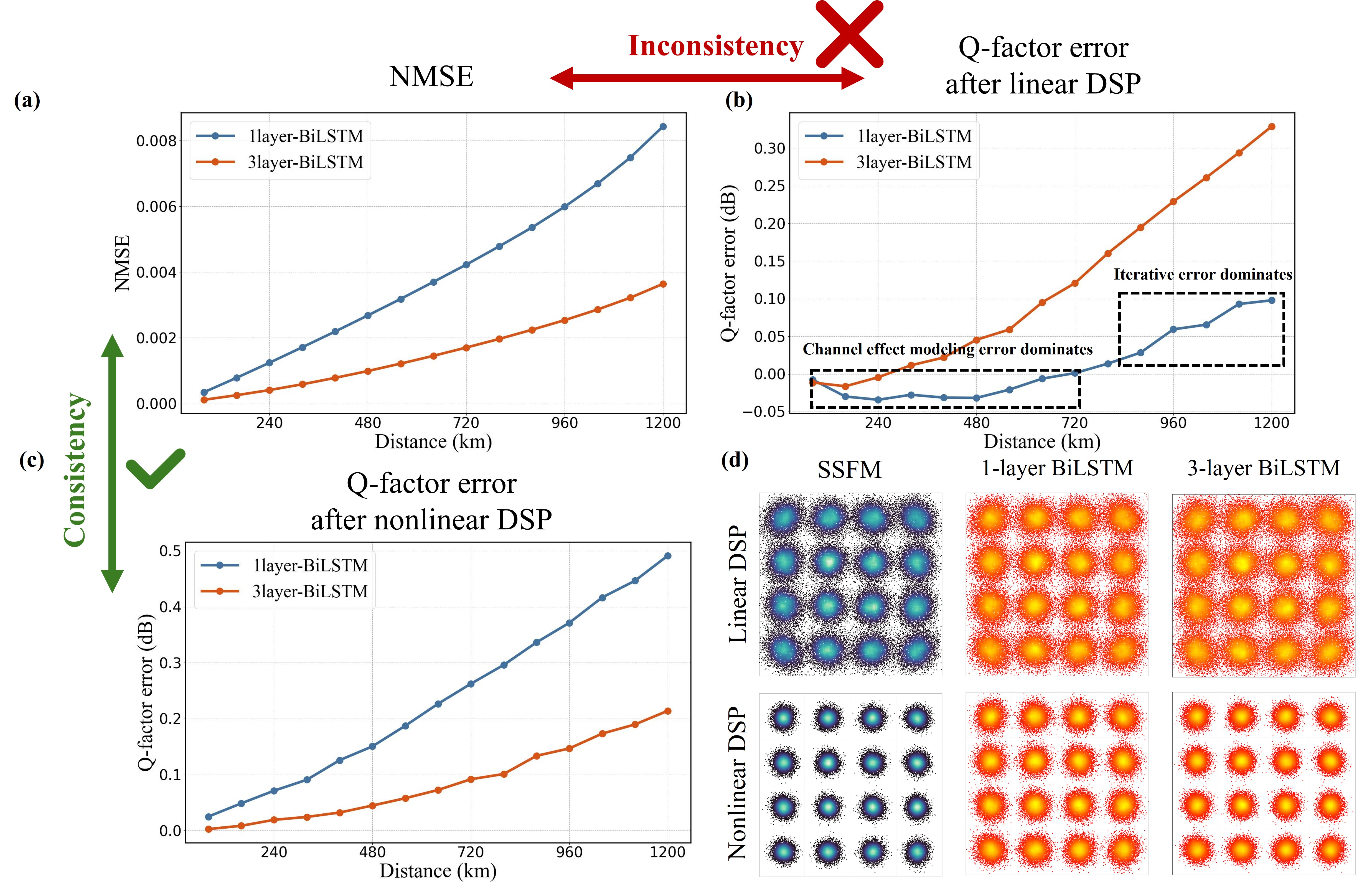}
\caption{Comparison of 1-layer and 3-layer BiLSTM between SSFM. (a) NMSE curve covering 1200 km. (b) Q-factor error after linear DSP processing covering 1200 km. (c) Q-factor error after nonlinear DSP processing covering 1200 km. (d) Constellations after linear and nonlinear DSP processing at 1200 km.} 
\label{fig4}
\end{figure*}

Here, we demonstrate the issue of the two types of errors canceling each other out and inconsistency between transmission performance and waveform errors through specific test results. We conduct a test using FDD-BiLSTM in a 3-channel WDM system with a transmission rate of 50 GBaud and a launch power of 4 dBm---approximately 2.5 dB higher than the optimal launch power. To adjust the accuracy of FDD-BiLSTM, we vary the number of layers in BiLSTM, which is used for nonlinear effects modeling. The number of layers was set to 1 and 3. Increasing the number of layers enhances the model's nonlinear fitting capacity, leading to higher model accuracy. Both models are trained on the same dataset with identical hyperparameters. First, waveform errors are evaluated using NMSE. Fig. \ref{fig4}(a) shows the NMSE curves over a 1200 km transmission range for both the 1-layer and 3-layer BiLSTM models. The 3-layer BiLSTM achieves consistently lower NMSE values than the 1-layer BiLSTM across a 1200 km transmission distance, demonstrating its superior waveform modeling capacity due to its deeper network and enhanced ability to capture nonlinearities. Next, transmission performance errors are assessed using the Q-factor after linear DSP. Fig. \ref{fig4}(b) shows that Q-factor errors for the 1-layer BiLSTM remain below 0.1 dB over the 1200 km transmission range, whereas Q-factor errors for the 3-layer BiLSTM increase with distance, peaking at 0.33 dB. These results indicate that the 1-layer BiLSTM more closely matches the transmission performance of SSFM, despite its inferior waveform modeling capacity, indicating the inconsistency between transmission performance and waveform errors. This makes it difficult for us to determine which DL model is the best. The fundamental reason for this phenomenon is that the channel effect modeling error and the iterative error cancel each other out under long-distance conditions. Fig. \ref{fig4}(b) illustrates the process by which the two types of errors cancel each other out over a long distance. At short transmission distances, the Q-factor errors for the 1-layer BiLSTM are negative, indicating the dominance of channel effect modeling errors, as iterative errors have not yet accumulated significantly. As the transmission distance increases, iterative errors accumulate and begin to dominate, degrading the Q-factor of the 1-layer BiLSTM and shifting its Q-factor errors into the positive region. At 1200 km, the combined effects of channel effect modeling errors and iterative errors result in a Q-factor error of only 0.1 dB for the 1-layer BiLSTM, but this does not indicate higher accuracy. In contrast, the Q-factor errors for the 3-layer BiLSTM remain positive throughout most of the transmission range, except for the initial spans. This occurs because the higher modeling capacity of the 3-layer BiLSTM reduces channel effect modeling errors, allowing iterative errors to dominate across the entire transmission distance. The dominance of iterative errors causes the Q-factor errors for the 3-layer BiLSTM to continually increase with transmission distance.

\add{In summary, due to the interference caused by the channel effect modeling errors and the iterative errors that cancel each other out after long-distance transmission, the Q-factor after linear DSP is difficult to accurately reflect the accuracy of the model. We need an additional metric to accurately evaluate the transmission performance of the DL model.}

\subsubsection{\textbf{DSP-assisted accuracy evaluation method}}

\begin{figure*}[!t]
\centering
\includegraphics[width=7in]{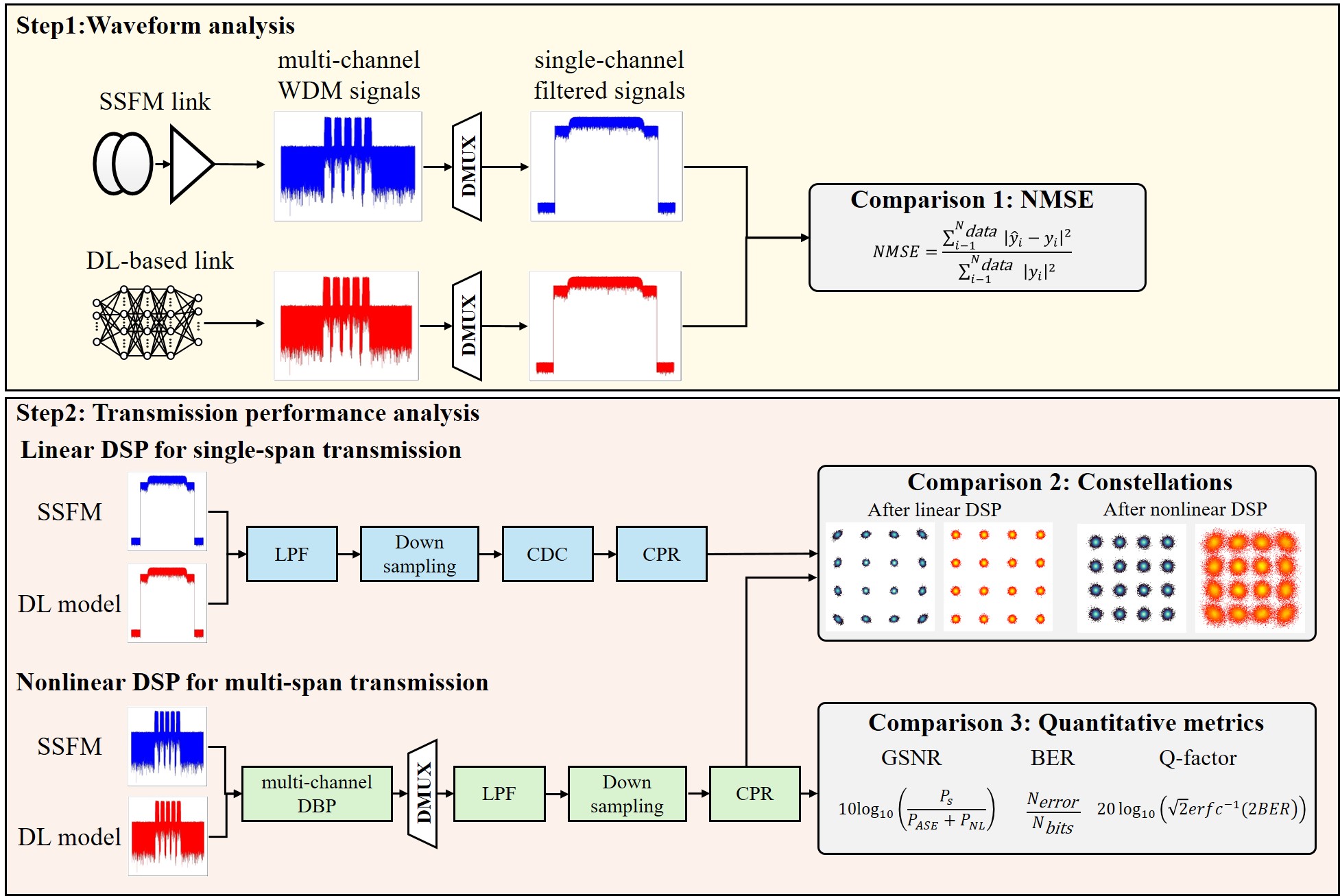}
\caption{The specific flow of DSP-assisted accuracy evaluation method.}
\label{fig5}
\end{figure*}

We have discussed the limitations of existing accuracy metrics, including the difficulty in defining a universally accepted NMSE threshold and the inconsistency between waveform errors and transmission performance errors after linear DSP. To address these limitations and establish a reliable, standardized method for comprehensively evaluating and comparing DL models, we propose a DSP-assisted accuracy evaluation method. This approach jointly considers both waveform errors and transmission performance errors and incorporates nonlinear DSP algorithms to eliminate the inconsistency between the two metrics. The DSP-assisted evaluation method consists of two stages: waveform analysis and transmission performance analysis, as illustrated in Fig. \ref{fig5}.

In the first stage, waveform analysis, full-field multi-channel WDM signals from both SSFM-based and DL-based optical fiber links are obtained and demultiplexed to extract CUT signals. The waveform contour in both the time and frequency domains provide an intuitive means for waveform analysis. Additionally, the NMSE of the the CUT signals is calculated to quantitatively assess waveform fidelity, excluding the influence of out-of-band noise. NMSE serves as an effective and relative metric for evaluating waveform modeling accuracy: lower NMSE values indicate higher fidelity in capturing the true channel response.

The second stage focuses on transmission performance analysis. Linear effects in the fiber channel can be accurately represented by physical models and effectively compensated using linear DSP. In contrast, the representation of nonlinear effects is more complex, making it the primary focus of DL modeling. Therefore, validating the model’s ability to capture nonlinear effects is critical. The accuracy of nonlinear modeling can be assessed by examining nonlinear phase rotations in the constellation after linear DSP. Furthermore, the ESNR severs as a quantitative metrics to evaluate the sparseness of the Rx constellation, which is influenced by accumulated nonlinear noise. However, after long-haul transmission, performance metrics after linear DSP may fluctuation due to partial cancellation between channel effect modeling error and iterative error, as previously analyzed. This phenomenon obscures the true modeling accuracy and leads to unreliable performance assessment. To address this limitation, we propose incorporating a nonlinear compensation algorithm---multi-channel DBP. This algorithm compensates for both linear and nonlinear impairments by inversely solving the NLSE. Unlike traditional single-channel DBP, multi-channel DBP operates on full-field WDM signals, enabling joint compensation of intra-channel and inter-channel nonlinearities. \add{The formula for multi-channel DBP can be expressed as:}
\begin{equation}
\begin{aligned}
\boldsymbol{A}(z-h, t) \approx & \exp \left(-\frac{h}{2} \hat{\boldsymbol{D}}\right) \exp \left\{-h \hat{\boldsymbol{N}}\left[\boldsymbol{A}\left(z-\frac{h}{2}, t\right)\right]\right\} \\
& \times \exp \left(-\frac{h}{2} \hat{\boldsymbol{D}}\right)\boldsymbol{A}(z, t),
\end{aligned}
\end{equation}
The parameters of multi-channel DBP are set to match the forward SSFM simulation: the nonlinear coefficient $\gamma$ is set to $1.3/(W \cdot km)$, and the step size is determined using the maximum phase rotation method with the maximum phase rotation of 0.005. This configuration ensures that the multi-channel DBP process is the exact inverse of the SSFM propagation, enabling complete compensation of nonlinear impairments. \add{It is worth noting that, since polarization-dependent effects are not included in our simulation system, DBP is not combined with a multiple-input multiple-output (MIMO) adaptive equalization filter. In practical experimental systems, MIMO equalization must be integrated with DBP to compensate for polarization-related impairments.} Incorporating multi-channel DBP offers two key advantages. First, the inverse compensation process enables a direct assessment of whether the forward nonlinear modeling by the DL model is consistent with the SSFM reference. Second, it mitigates the performance fluctuations caused by error cancellation after linear DSP. As illustrated in Fig. \ref{two_errors}(a). DL models often exhibit underfitting of nonlinear effects compared to SSFM, leading to compensation overflow during full inverse DBP. \add{Since DBP operates by solving the NLSE backward, any unmodeled nonlinearity in the forward process manifests as excess nonlinear distortion in the reverse process. These additional impairments resemble the signature of underfitting—except that the nonlinear phase rotation occurs in the opposite direction.} After applying multi-channel DBP, forward modeling errors are transformed into residual impairments that degrade transmission performance in a manner similar to iterative errors. This mechanism eliminates the misleading cancellation effect, stabilizes the evaluation outcome, and ensures a more accurate and fair comparison of DL models.

To validate the effectiveness of the multi-channel DBP algorithm, we compare the performance of the SSFM reference, a 1-layer BiLSTM, and a 3-layer BiLSTM model. Fig. \ref{fig4}(d) shows constellations after both linear and nonlinear DSP at an 800 km transmission distance. After linear DSP, constellations of both BiLSTM models closely resemble that of SSFM, making their performance difficult to distinguish. However, after nonlinear DSP, the 3-layer BiLSTM constellation aligns more closely with SSFM, while the 1-layer BiLSTM exhibits significant deviations and pronounced nonlinear distortions. These observations confirm that multi-channel DBP effectively transforms insufficient forward nonlinear modeling into additional nonlinear impairments during inverse compensation. Further validation is provided by analyzing Q-factor errors. Fig. \ref{fig4}(c) illustrates Q-factor errors after nonlinear DSP over a 1200 km transmission range. The 1-layer BiLSTM exhibits higher Q-factor errors, peaking at 0.5 dB, compared to the 3-layer BiLSTM, which has a maximum error of only 0.21 dB. These results indicate that the 3-layer BiLSTM more accurately captures the nonlinear effects as modeled by SSFM. As shown in Fig. \ref{fig4}(a) and (c), the results of NMSE and Q-factor errors after nonlinear DSP are consistent across the transmission distance. This alignment demonstrates that multi-channel DBP effectively mitigates the performance fluctuations caused by error cancellation after linear DSP and establishes consistency between waveform errors and transmission performance errors over long-haul transmission. 

\add{The proposed DSP-assisted accuracy evaluation method integrates both waveform analysis and transmission performance evaluation. By incorporating multi-channel DBP, it resolves the issue of unreliable performance assessment after linear DSP and ensures consistency between waveform errors and transmission performance errors. As a result, it provides a fair and comprehensive benchmark for comparing various DL models, which is applied in the subsequent section.}

\section{Comparison of DL-based schemes}
DL-based optical fiber channel waveform modeling schemes have demonstrated comparable accuracy with significantly reduced complexity compared to traditional SSFM. In Section III, we have introduced different types of DL schemes and their training methodologies. Currently, DL models are primarily applied to single-channel and few-channel WDM systems, or simplified wideband WDM configurations, with their performance in wideband scenarios remaining inadequately explored. Additionally, there is a lack of fair and comprehensive comparison of DL models under unified conditions and standards. In this section, we present a comprehensive comparison of different types of DL schemes based on the DSP-assisted accuracy evaluation methods proposed in Section IV. To provide a more thorough analysis of their performance, the comparisons begin with simpler scenarios, progressing from few-channel and low-rate to multi-channel and high-rate wideband configurations. By selecting progressively more challenging scenarios, we aim to identify the best-performing schemes and reveal their application potential in wideband environments. Furthermore, based on results exhibited in wideband scenarios, we analyze the challenges faced by DL schemes, considering both more intricate linear and nonlinear effects, as well as higher sampling rates. This analysis provides valuable insights that will inform the future optimization of DL models for improved performance in wideband channel waveform modeling.

\subsection{Comparisons in few-channel and low-rate WDM systems}

DL schemes have demonstrated outstanding performance in few-channel WDM systems. Therefore, the first scenario for comparison is set to a 5-channel WDM configuration. The transmission rate is set at 50 GBaud, with a launch power of 4 dBm per channel, creating a highly nonlinear scenario that demands stronger nonlinear modeling capacity of DL models. To provide a thorough comparison of DL approaches, the comparison is divided into two parts based on the classification of schemes in Section III, including overall versus distributed methods and pure-data driven versus data-physics hybrid-driven methods. 

\subsubsection{\textbf{Overall schemes vs. distributed schemes}}

\begin{table*}[!ht]
    \centering
    \renewcommand\arraystretch{1.6}
    \caption{The parameters of DL models for comparison between overall and distributed schemes.}
    \begin{tabular}{cccccc}
    \hline\hline
        ~ & \multicolumn{2}{c}{Overall-CGAN} & \multirow{2}{*}{~} & \multirow{2}{*}{Overall-BiLSTM} & \multirow{2}{*}{Distributed-BiLSTM}\\ 
        ~ & Generator & Discriminator & ~ & ~ &\\ 
        \hline
        Input size & 13200 & 2 & Input size & 80 & 80 \\ 
        Hidden layer & 4 & 4 & Layers & 3 & 3\\ 
        Hidden size & [52600,6575,3287,1643,80] & [52640,6580,3290,1645,1] & Hidden size & 80 & 80 \\ 
        Output size & 80 & 20 & Time steps & 657 & 165 \\ 
        \hline\hline
    \end{tabular}
    \label{overall network parameter}
\end{table*}

\begin{figure}[!t]
\centering
\includegraphics[width=3in]{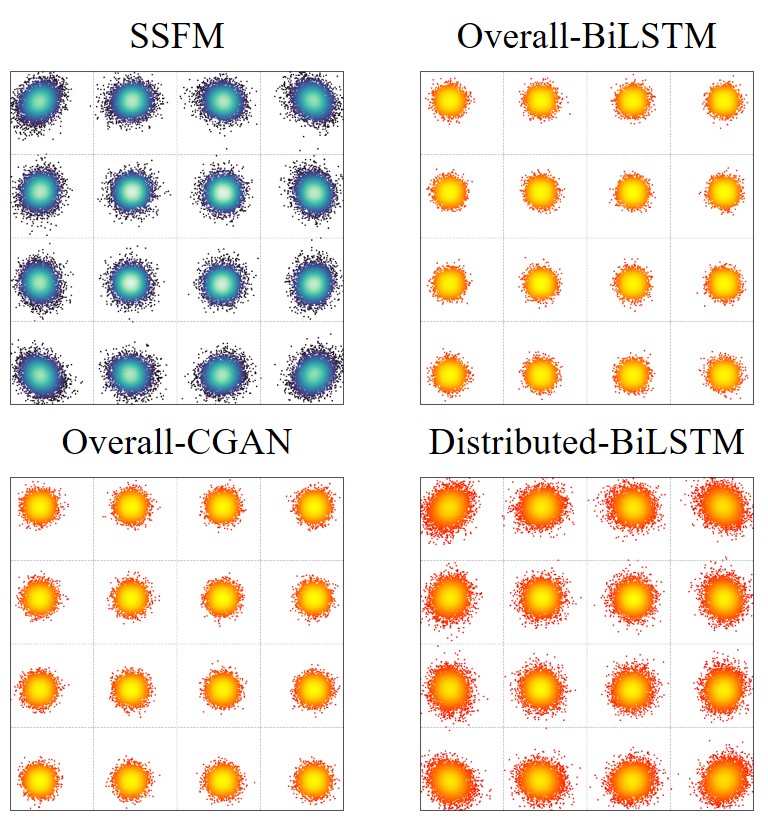}
\caption{Constellations of SSFM, overall-CGAN, overall-BiLSTM and distributed-BiLSTM at 320 km transmission.}
\label{fig6}
\end{figure}

\begin{table}
    \centering
    \renewcommand\arraystretch{1.5}
    \caption{Performance comparison of overall and distributed schemes at 800 km transmission.}
    \begin{tabular}{ccc}
    \hline\hline
    {Scheme} & {NMSE} & {Q-factor error(dB)} \\
    \hline
    {Overall-BiLSTM} & 0.0133 & 0.53 \\
    {Overall-CGAN} & 0.0151 & 0.67 \\
    {Distributed-BiLSTM} & 0.0011 & 0.16 \\
    \hline\hline
    \end{tabular}
    \label{comparison_table}
\end{table}

We compare the performance of both overall and distributed schemes. Specifically, the models compared include overall-BiLSTM \cite{BiLSTM}, overall-CGAN \cite{CGAN}, and distributed-BiLSTM. The parameters of these DL models are listed in Table \ref{overall network parameter}. The overall schemes utilize a single network to model the entire 320 km fiber link, \add{while the distributed scheme cascades four models with identical parameters to handle each 80 km fiber span separately.} To address the ISI effects caused by CD accumulation over distances, the input window length for distributed schemes is set at 211 symbols, while for overall schemes, it is 4 times that of the distributed schemes (844 symbols) due to longer CD accumulation.

To compare the performance between distributed schemes and overall schemes, we first examine their waveform modeling capabilities. The NMSE for each scheme at a transmission distance of 320 km is presented in Table \ref{comparison_table}. Here, the NMSE is 0.0133 for overall-BiLSTM and 0.0151 for overall-CGAN, whereas the distributed-BiLSTM achieves a significantly lower NMSE of 0.0011. This indicates a more precise waveform modeling capability of the distributed schemes. Further analysis of transmission performance is conducted by examining the constellations after linear DSP, as shown in Fig. \ref{fig6}. The shapes of the constellations from overall schemes are dissimilar to that of SSFM. Additionally, the Q-factor error after nonlinear DSP for overall-BiLSTM is 0.53 dB and 0.67 dB for overall-CGAN, whereas while that of the distributed-BiLSTM is only 0.16 dB, indicating a significant performance gap between overall schemes and SSFM.

The excellent results in waveform errors and transmission performance errors for distributed-BiLSTM highlight the better modeling capacity of distributed schemes. The BiLSTM is a deterministic model, which is inherently limited in handling random features. This limitation makes overall-BiLSTM particularly unsuitable for modeling long-haul fiber links with random noise between each fiber span. \add{Furthermore, the adversarial training process of CGAN makes it difficult for it to fully converge in multi-channel WDM systems} \cite{GAN_challenge, gan_review}. In contrast, distributed schemes, which individually model the channel effects of each span, simplify the linear and nonlinear effects accumulated with the transmission distance, thus enhancing accuracy. Meanwhile, the random noise between each span can be directly modeled by a Gaussian distribution, eliminating the need for DL models to learn these random features. Therefore, distributed schemes are highly suitable for long-distance, multi-channel WDM systems and represent the main technical approach for future research.


\begin{table*}[!ht]
    \centering
    \renewcommand\arraystretch{1.6}
    \caption{\add{The parameters of DL models for comparison between pure data-driven and data-physic hybrid-driven schemes.}}
    \begin{tabular}{c ccccc c} 
    \hline\hline
        ~ & ~ & \multicolumn{2}{c}{DeepONet} & ~ & FNO & ~ \\
        ~ & ~ & Branch Net & Trunk Net & ~ & ~ & ~  \\
        \hline
        ~ & Input size & 13200 & 2 & Input size & 13200$\times$1 & ~  \\
        ~ & Hidden layer & 4 & 4 & Fourier layer & 4 & ~ \\
        ~ & Hidden size & {[}6600,3300,1650,825,412{]} & {[}256,256,256,256,256{]} & Hidden dimension & 16 & ~ \\
        ~ & Output size & 80 & 20 & Fourier mode & 6600 & ~ \\
        \hline\hline
        ~ & BiLSTM & FDD-BiLSTM & Seq2Seq-FDD & ~ & Transformer & FDD-Transformer \\
        \hline
        Input size & 80 & 80 & 80 & Input size & 80 & 80 \\
        Hidden size & 80 & 80 & 80 & Hidden size/FFN size & 80/320 & 80/320 \\
        Layers & 3 & 3 & 3 & Layers & 3 & 3 \\
        Time step & 165 & 45 & 544 & Time step & 165 & 45 \\
        \hline\hline
    \end{tabular}
    \label{network parameter}
\end{table*}

\begin{figure*}[!t]
\centering
\includegraphics[width=7in]{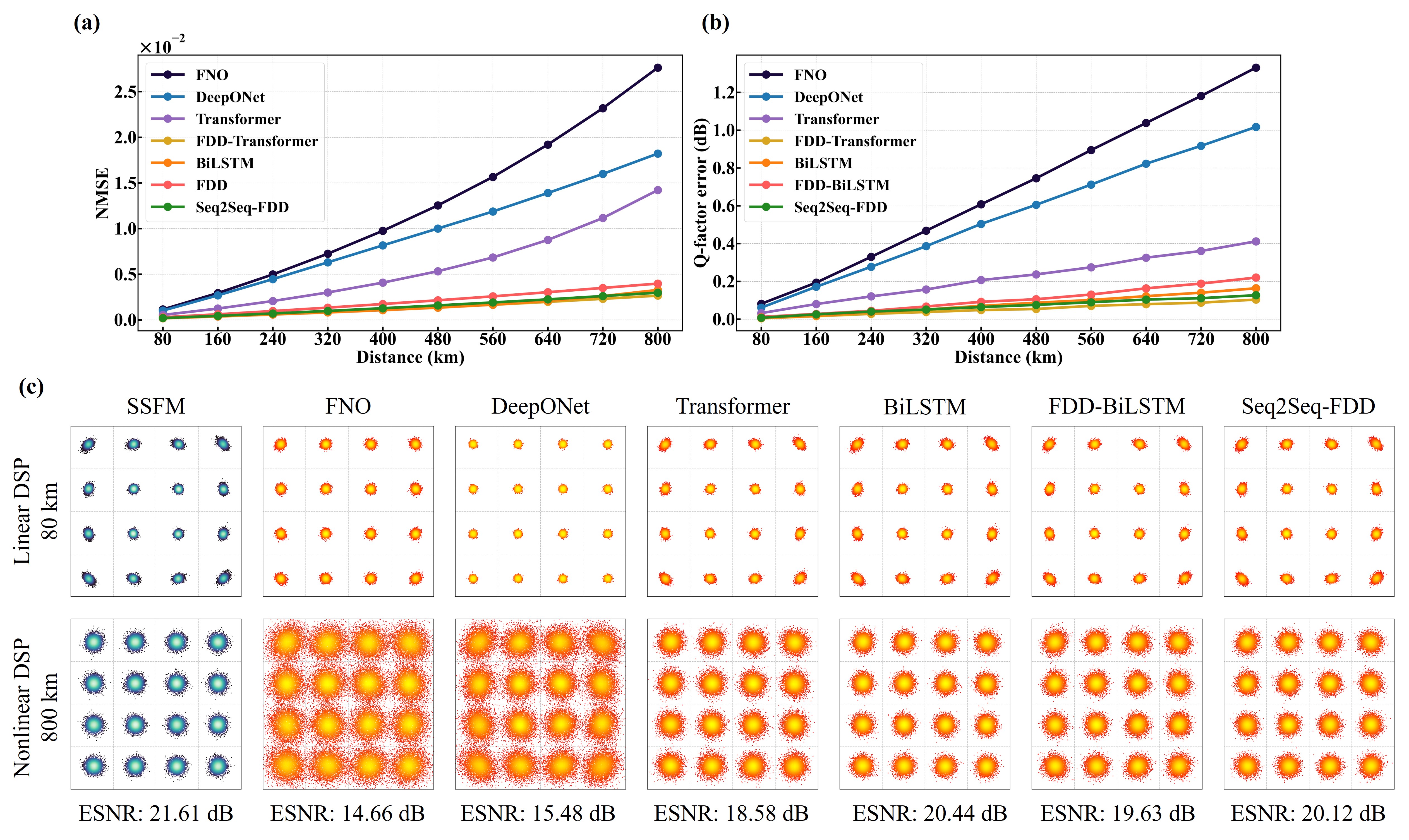}
\caption{\add{Comparison of SSFM and DL-based schemes, including pure data-driven schemes and data-physic hybrid-driven schemes. (a) NMSE curve of DL-based schemes covering 800 km. (b) Q-factor error curve after nonlinear DPS processing between SSFM and DL-based schemes. (c) Constellations after linear DSP processing of SSFM and DL-based schemes at 80 km and 800km.}}
\label{fig7}
\end{figure*}

\subsubsection{\textbf{Pure data-driven schemes vs. data-physic hybrid-driven schemes}}
Distributed schemes offer significant advantages in accuracy and flexibility in handling random noise, making them the preferred design in recent DL approaches. Therefore, the next comparison employs only distributed schemes. This section contrasts pure data-driven methods and data-physics hybrid-driven methods, involving both neural networks and neural operators. The pure data-driven schemes include BiLSTM, Transformer, FNO, and DeepONet, while the data-physics hybrid-driven scheme is represented by FDD-BiLSTM, \add{FDD-Transformer and Seq2Seq-FDD. In pure data-driven schemes, DL models are responsible for capturing both linear and nonlinear effects, whereas in FDD-based hybrid schemes, they focus primarily on modeling nonlinear effects due to the decoupling of linearity.} For accurately modeling linear effects, the pure data-driven schemes require a longer input window to account for ISI, while the input window length of FDD-BiLSTM is reduced because the nonlinear inter-symbol correlations are shortened, as determined by equations (\ref{Non_ISI1}) and (\ref{Non_ISI2}). Therefore, pure data-driven schemes utilize an input window of 165 symbols (82+1+82), whereas the FDD-BiLSTM and FDD-Transformer requires only 45 symbols (22+1+22). \add{The Seq2Seq-FDD employs a multi-symbol output mode during inference, with an input window of 544 symbols (22+500+200).} Other model parameters are summarized in Table \ref{network parameter}.

\add{To evaluate the long-haul transmission performance, each DL model is tested over 10 propagation iterations for an 800 km link and compared to SSFM. Waveform errors are quantified using the NMSE. As shown in Fig.} \ref{fig7}\add{(a), BiLSTM, FDD-BiLSTM, FDD-Transformer and Seq2Seq-FDD exhibit excellent waveform modeling accuracy, with NMSE values below 5E-3 over the entire 800 km transmission. The NMSE values for BiLSTM, FDD-BiLSTM and Seq2Seq-FDD are 3.3E-3, 4E-3 and 3E-3, respectively. Although slight differences exist, the performance gap is minimal. FDD-Transformer achieves the highest accuracy, yielding an NMSE of 2.7E-3 at 800km. In contrast, Transformer performs moderately worse, with NMSE values just below 1.5E-2. FNO and DeepONet exhibit significantly larger discrepancies, with NMSE values of 2.8E-2 and 1.8E-2 at 800 km, respectively. Notably, the NMSE of FNO is approximately ten times higher than that of FDD-Transformer at 800 km, indicating substantial waveform distortion. The overall waveform modeling accuracy ranks as follows:} $ \text{FDD-Transformer} < \text{Seq2Seq-FDD} < \text{BiLSTM} < \text{FDD-BiLSTM} < \text{Transformer} < \text{DeepONet} < \text{FNO}$.

\add{Next, we evaluate transmission performance using constellations, ESNR and Q-factor errors. To analyze single-span performance, the first row of Fig.} \ref{fig7}\add{(c) presents the constellations of the DL models and SSFM after linear DSP at 80 km transmission. The SSFM constellation shows clear nonlinear phase rotation due to the high launch power (4.0 dBm). The constellations of Transformer, BiLSTM, FDD-BiLSTM, and Seq2Seq-FDD display similar nonlinear phase rotations and closely resemble the SSFM. In contrast, FNO and DeepONet exhibit Gaussian-like distributions with negligible phase rotation, indicating poor capture of nonlinear effects. To further evaluate long-haul performance, the second row of Fig.} \ref{fig7}\add{(c) exhibits the constellations and ESNR after nonlinear DSP at 800 km. Multi-channel DBP effectively compensates for nonlinear impairments in the SSFM signal, resulting in an ESNR of 21.61 dB. The constellations of Transformer, BiLSTM, FDD-BiLSTM, and Seq2Seq-FDD exhibit comparable compensation quality. Among them, Transformer exhibits slightly elevated nonlinear noise due to underfitting, leading to an ESNR degradation to 18.58 dB. In contrast, FNO and DeepONet introduce significant additional noise when combined with multi-channel DBP, resulting in severe ESNR degradation to 14.66 dB and 15.48 dB, respectively—indicating very poor nonlinear modeling accuracy. Finally, Q-factor errors are employed to quantitatively assess transmission performance, as presented in Fig.} \ref{fig7}\add{(b). Q-factor errors for all DL models increase with transmission distance, consistent with NMSE trends, due to cumulative modeling errors. BiLSTM, FDD-BiLSTM, FDD-Transformer, and Seq2Seq-FDD show nearly identical Q-factor errors, with a maximum error of approximately 0.22 dB after 800 km, indicating excellent agreement with SSFM. Transformer performs slightly worse, reaching a maximum Q-factor error of 0.41 dB. FNO and DeepONet show rapid error growth, reaching 1.33 dB and 1.02 dB, respectively, revealing a substantial performance gap relative to the SSFM reference.}

\add{Results from NMSE, constellations and Q-factor errors show strong consistency, validating the effectiveness and fairness of the DSP-assisted accuracy evaluation method. The accuracy ranking of DL schemes, based on both NMSE and Q-factor metrics, is as follows:} $ \text{FDD-Transformer} < \text{Seq2Seq-FDD} < \text{BiLSTM} < \text{FDD-BiLSTM} < \text{Transformer} < \text{DeepONet} < \text{FNO}$. \add{This ranking reveals that high-accuracy DL models generally share two key characteristics: (1) the adoption of a temporal neural network architecture, and (2) integration with the FDD scheme. First, the temporal neural networks—such as BiLSTM and Transformer—leverage their inherent temporal modeling capabilities (via recurrent connections or self-attention mechanisms) to effectively capture inter-symbol correlations induced by CD. In contrast, FNO and DeepONet rely primarily on fully connected and convolution structures, which are less effective at modeling long-term temporal dependencies. While these models perform well in single-channel scenarios, their accuracy degrades significantly in the 5-channel WDM system, where stronger inter-symbol correlations increase the modeling challenge. Moreover, the fully connected architecture typically processes one-dimensional input. In the 5-channel case, an input window of 165 symbols results in a flattened input vector of dimension 13,200. Such a high-dimensional input poses significant challenges for FNO and DeepONet in terms of both computational complexity and feature learning efficiency. Second, beyond the advantages of temporal modeling, the FDD framework contributes significantly to improved performance. By using a physical model to account for linear effects, FDD allows the DL models to focus exclusively on nonlinear effects. This design reduces the complexity of the learning task, shortens the length of inter-symbol correlations. In summary, temporal neural networks and the FDD architecture are better suited for accurate modeling in multi-channel optical transmission scenarios.}

\add{Here, we primarily focus on evaluating the accuracy of the DL model. However, generalization capability and inference time are also critical metrics for assessing the practicality of the approach. Due to space constraints, we provide a brief discussion of the model’s generalization ability and inference time in Appendices B and C, respectively.}

\begin{figure}[!t]
\centering
\includegraphics[width=3.3in]{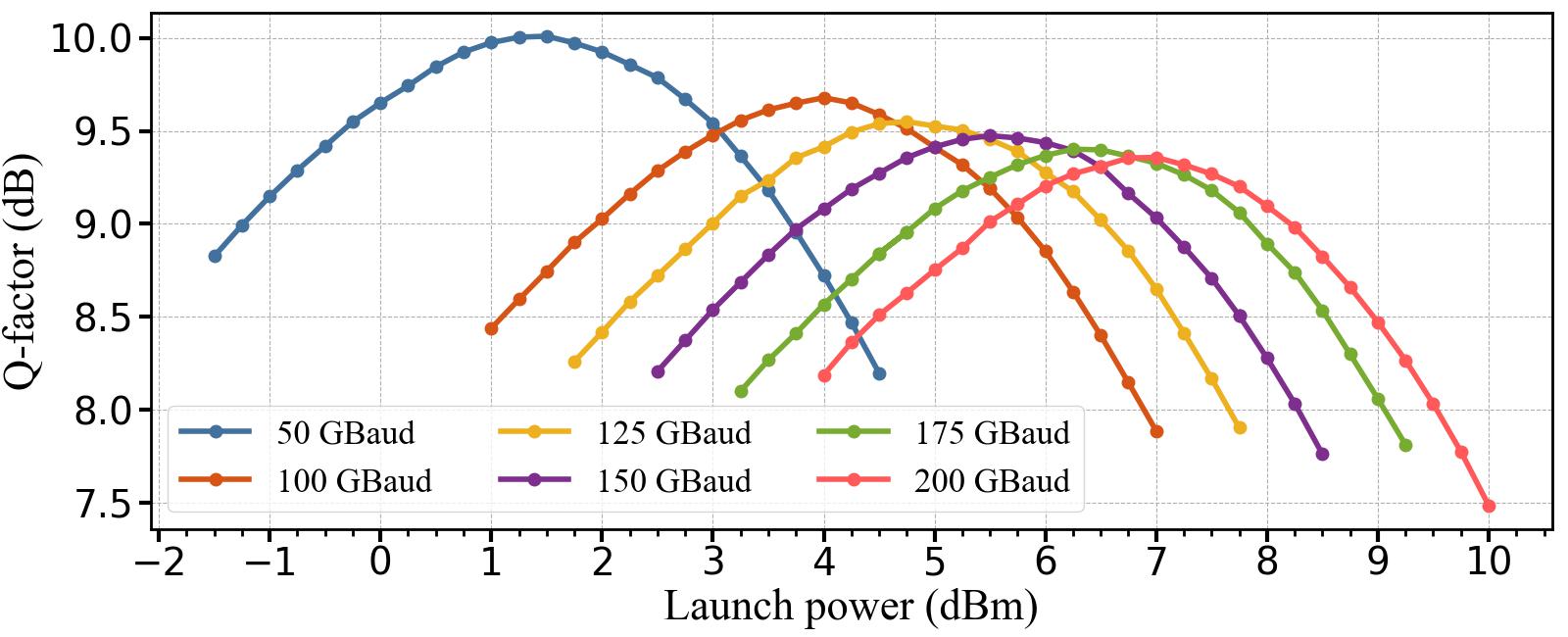}
\caption{Launch powers versus Q-factor at various symbol rates.}
\label{fig8}
\end{figure}

\begin{table}
    \centering
    \renewcommand\arraystretch{1.5}
    \caption{launch powers at various symbol rates}
    \label{optim_power}
    \begin{tabular}{ccc}
    \hline\hline
    {Symbol rate} & {Optimal launch power} & {Testing launch power} \\
    {(GBaud)} & {(dBm)} & {(dBm)} \\
    \hline
    50 & 1.5 & 4.0 \\
    100 & 4.0 & 6.5 \\
    125 & 4.75 & 7.25 \\
    150 & 5.5 & 8.0 \\
    175 & 6.25 & 8.75 \\
    200 & 7.0 & 9.5\\
    \hline\hline
    \end{tabular}
\end{table}


\subsection{Comparisons in more-channel and higher-rate WDM systems.}
\begin{figure*}[!t]
\centering
\includegraphics[width=7in]{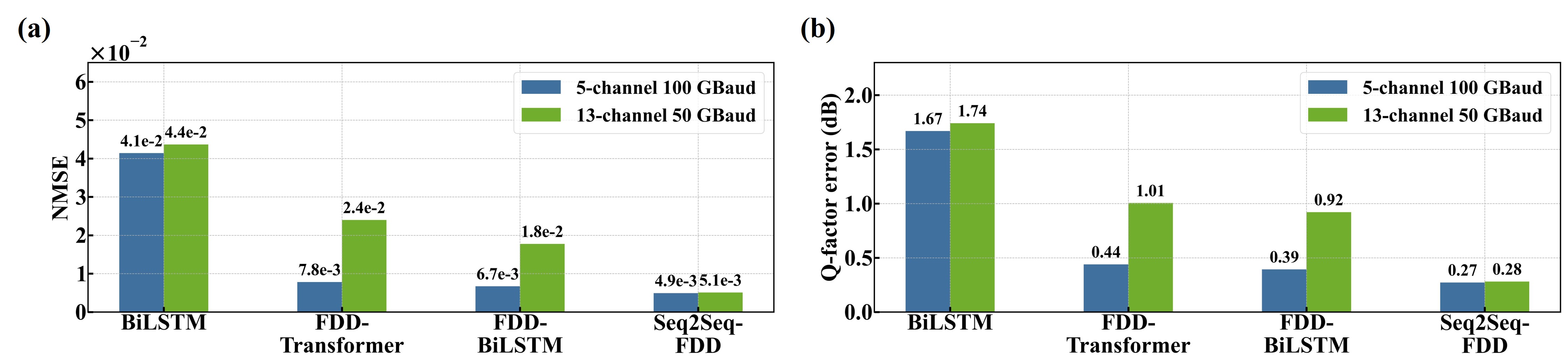}
\caption{\add{Comparison results under 13-channel with 50 GBaud and 5-channel with 100 GBaud after 800 km transmission. (a) NMSE. (b) Q-factor errors.}}
\label{fig9}
\end{figure*}

In a 5-channel WDM configuration, the temporal neural networks demonstrate excellent modeling accuracy. To further evaluate their performance under more demanding conditions, we extend the assessment to more complex WDM scenarios with more channel number and higher symbol rates. The test cases include: (1) a 13-channel WDM system operating at 50 GBaud with a launch power of 4.0 dBm, and (2) a 5-channel WDM system operating at 100 GBaud with a launch power of 6.5 dBm. The launch power in the 100 GBaud case is increased to 6.5 dBm to align with the higher optimal launch power associated with the elevated symbol rate. The trend of optimal launch power as a function of symbol rate is illustrated in Fig. \ref{fig8}, \add{based on SSFM simulations.} The specific values of optimal and test launch powers for different symbol rates are summarized in Table \ref{optim_power}. In these scenarios, the length of ISI and nonlinear inter-symbol correlations increases due to the wider spectrum and higher symbol rate. Consequently, the input window length is set to 427 symbols for non-FDD schemes and 117 symbols for FDD-based schemes in the 13-channel, 50 GBaud case, and to 601 and 163 symbols, respectively, in the 5-channel, 100 GBaud case. Due to the significantly longer input windows, the memory requirements for FNO, DeepONet, and Transformer exceed the capacity of our training infrastructure. Moreover, these models already exhibited degraded accuracy in the 5-channel, 50 GBaud scenario. Therefore, in this section, we exclude FNO, DeepONet, and Transformer from the comparative analysis.

All DL models are evaluated over 10 propagation iterations for an 800 km transmission distance. The NMSE and Q-factor errors at 800 km are shown in Fig. \ref{fig9}. In both WDM configurations, BiLSTM exhibits significant accuracy degradation, with Q-factor errors reaching 1.74 dB (13-channel, 50 GBaud) and 1.67 dB (5-channel, 100 GBaud), respectively. In contrast, FDD-based schemes maintain Q-factor errors generally below 1 dB, demonstrating superior performance. This improvement is attributed to the FDD framework, which reduces the input window length by decoupling linear effects via a physical model, allowing the DL component to focus on modeling nonlinear distortions. \add{Among the FDD-based models, FDD-BiLSTM slightly outperforms FDD-Transformer at 100 GBaud, with Q-factor errors of 0.39 dB and 0.92 dB, respectively, in the 100 GBaud and 13-channel scenarios. However, Seq2Seq-FDD achieves the highest accuracy, with Q-factor errors of only 0.27 dB and 0.28 dB, exhibiting improvements of 0.64 dB and 0.12 dB over FDD-BiLSTM. Although FDD-BiLSTM and FDD-Transformer adopt the FDD framework, their input windows remain relatively long (163 and 117 symbols, respectively), requiring the model to process over 100 time steps per inference. This poses challenges for training convergence. In contrast, Seq2Seq-FDD introduces a multi-input multi-output  architecture combined with transfer learning. In the first training stage, the model processes shorter input windows, prioritizing the learning of nonlinear features from the most adjacent symbols—which have the strongest influence—resulting in a smaller effective model scale and faster convergence. In the second stage, the model is fine-tuned on the full input window, leveraging the learned features from the first stage. This two-stage strategy enables stable convergence even with large input windows. These architectural and training advantages make Seq2Seq-FDD particularly well-suited for wideband WDM channel modeling.}

\subsection{Extending to wideband WDM systems}
\begin{figure*}[!t]
\centering
\includegraphics[width=7in]{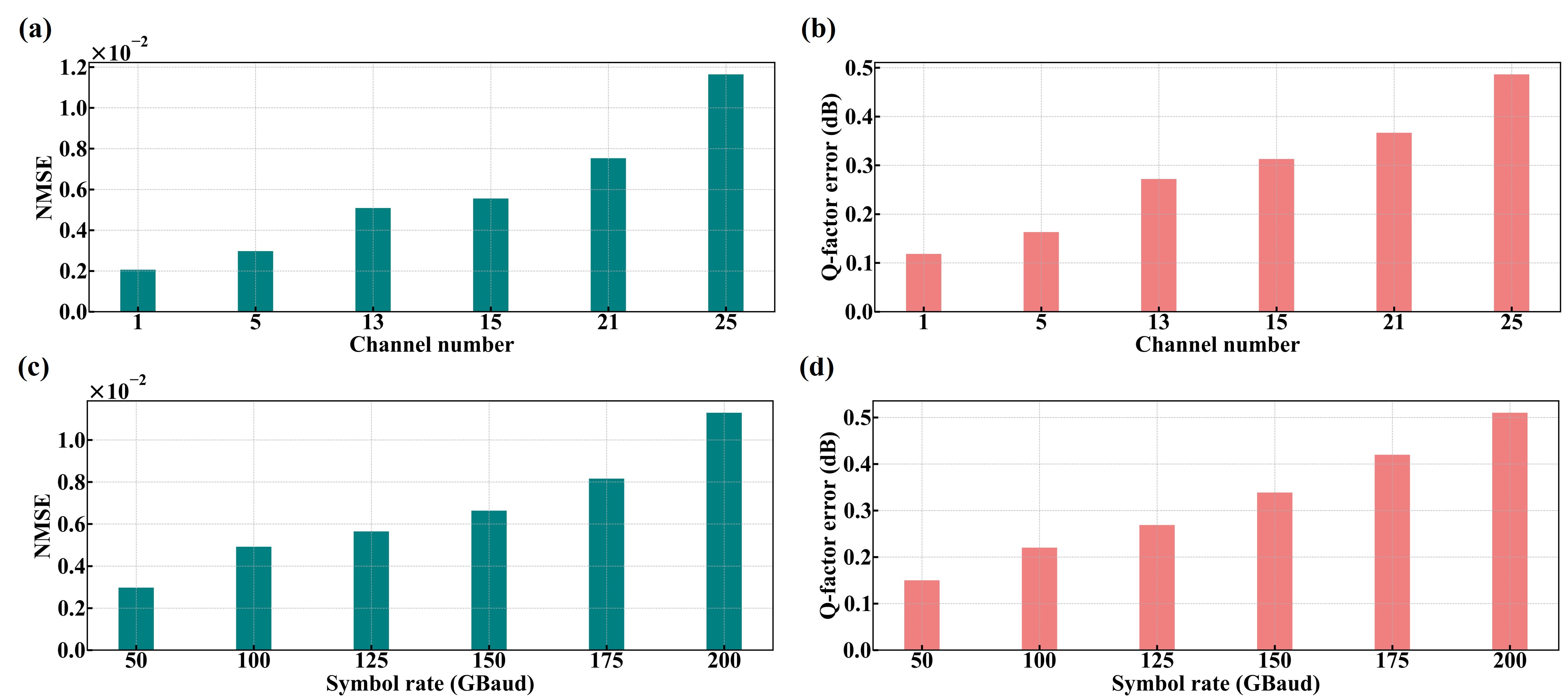}
\caption{\add{The results of Seq2Seq applied in more-channel WDM systems after 800 km transmission. (a) NMSE across different number of channels. (b) Q-factor across different symbol rates. (c) NMSE across different number of channels. (d) Q-factor across different symbol rates.}}
\label{fig10}
\end{figure*}


\add{In the previous comparison, Seq2Seq-FDD has demonstrated excellent modeling capabilities. However, accuracy degradation is observed in WDM configurations with more channels and higher rates. To better analyze its effectiveness in wideband WDM systems, we extend Seq2Seq-FDD to WDM configurations with up to 25 channels and a symbol rate of 200 GBaud.} Tests are conducted across channel configurations of [1, 5, 13, 15, 21, 25] channels, with a fixed transmission rate of 50 GBaud and a launch power of 4 dBm per channel. Another test is conducted using symbol rate configurations of [50, 100, 125, 150, 175, 200] GBaud, with the number of channels fixed at 5. The optimal launch power is increased for higher symbol rates. The launch power for testing is adjusted to be approximately 2.5 dB higher than the optimized launch power, as detailed in the Table \ref{optim_power}. 

\add{We first evaluate the performance under more-channel configuration. The NMSE and Q-factor errors performance of Seq2Seq-FDD across various channel configurations at 800 km is shown in Fig.} \ref{fig10}\add{(a) and (b). At 800 km transmission, the NMSE is just 2.97E-3 for the 5-channel configuration, compared to 7.53E-3 for 21-channel and 1.16E-2 for 25-channel configurations. This indicates a gradual decline in waveform modeling capacity as the number of channels increases. For the 5-channel scenario, the Q-factor error is just 0.16 dB at 800 km transmission, compared to 0.49 dB in the 25-channel configuration, indicating a significant transmission performance gap in more-channel configurations. The longer inter-symbol correlations, more intricate nonlinear representations, and higher sampling rate of SSFM are the main reasons for this performance degradation. }

\add{Next, we evaluate the performance under higher-rate configuration. he NMSE and Q-factor errors performance of Seq2Seq-FDD across various symbol rate configurations over an 800 km transmission is shown in Fig.} \ref{fig10}\add{(c) and (d). After 800 km transmission, the NMSE increases with symbol rates, which is just 2.97E-3 for 50 GBaud configuration, compared to 8.16E-3 for 175 GBaud and 1.13E-2 for 200 GBaud configurations, indicating significant waveform errors in higher-rate scenarios. For the 50 GBaud scenario, the Q-factor error is just 0.15 dB at 800 km transmission, compared to 0.51 dB for 200 GBaud configuration, indicating a significant transmission performance gap in higher-rate configurations. The combined effects of longer inter-symbol correlation and stronger nonlinear distortions due to higher launch powers jointly contribute to the degradation in modeling accuracy.}

\subsection{Analysis of challenges for the application of DL schemes in wideband WDM systems}

The results from NMSE and Q-factor errors collectively reveal accuracy degradation for Seq2Seq-FDD as the number of channels and symbol rates increase. Here, we analyze the reasons of this accuracy degradation from three perspectives: the more intricate linear effects and nonlinear effects, as well as the higher sampling rate of the SSFM in the digital simulation system.

Linear effects, primarily CD, induce ISI. The number of symbols affected by ISI is determined by the spectral width and symbol rate of the WDM signals, as described by equation (\ref{ISI}). \add{In more-channel configurations with fixed symbol rate, increasing the number of channels broadens the overall spectrum, thereby extending the ISI length. Similarly, in high-rate configurations with a fixed channel number, a higher symbol rate increases both the symbol rate and spectral width simultaneously, leading to a quadratic increase in ISI. As a result, the model requires a longer input window to accurately capture the temporal correlations introduced by ISI—particularly for non-FDD models, which must learn both linear and nonlinear effects from data. Temporal neural networks—such as BiLSTM and Transformer—excel at modeling temporal dependencies through their recurrent structures and self-attention mechanisms, enabling more effective learning of long ISI. Consequently, they achieve higher accuracy compared to models based on fully connected architectures. Furthermore, the FDD framework mitigates this challenge by decoupling linear effects via a physical model, allowing the DL component to focus on shorter nonlinear inter-symbol correlations. However, in multi-channel and high-rate systems, even FDD-based models must capture extended nonlinear inter-symbol correlations, which remains challenging for capturing long-term memory. As a result, model accuracy degrades in wideband scenarios.}

Nonlinear effects, primarily referring to Kerr nonlinearity, can be categorized into intra-channel and inter-channel nonlinearities, as described in equation \ref{nonlinear equ}. The overall strength of nonlinear effects is influenced by the signal power. The inter-channel nonlinearities also become more pronounced as the number of channel increases. Consequently, as the number of channels grows, inter-channel nonlinear effects increase significantly, complicating the ability of a single DL model to accurately capture these nonlinearities across all channels. Moreover, as the symbol rate increases, the optimal launch power also rises, as illustrated in Fig. \ref{fig8}, which further challenges the DL model by introducing stronger nonlinear effects due to higher launch powers. Therefore, the escalation of nonlinear effects, particularly the growing complexity of inter-channel nonlinearities, presents a key challenge for waveform modeling in wideband WDM systems. To address these challenges, DL models must possess enhanced nonlinear fitting capabilities to effectively model the stronger and more intricate nonlinearities encountered in these systems.

In digital simulation systems, the sampling rate of the SSFM is typically set to four times the number of WDM channels. This choice aims to avoid spectral aliasing in the full-field WDM signal, enable the construction of analog-like waveforms, and strike a balance between simulation accuracy and computational complexity. In multi-channel configurations, the increased sampling rate results in more sampling points per symbol, which expands the input dimension and parameter scale of the DL model. While larger-scale models have the potential to achieve higher accuracy, they are more challenging to train to convergence, requiring larger datasets, longer training times, and more sophisticated training strategies.

\section{Potential solutions for enhancing the accuracy of DL models}

\begin{figure}[!t]
\centering
\includegraphics[width=3.5in]{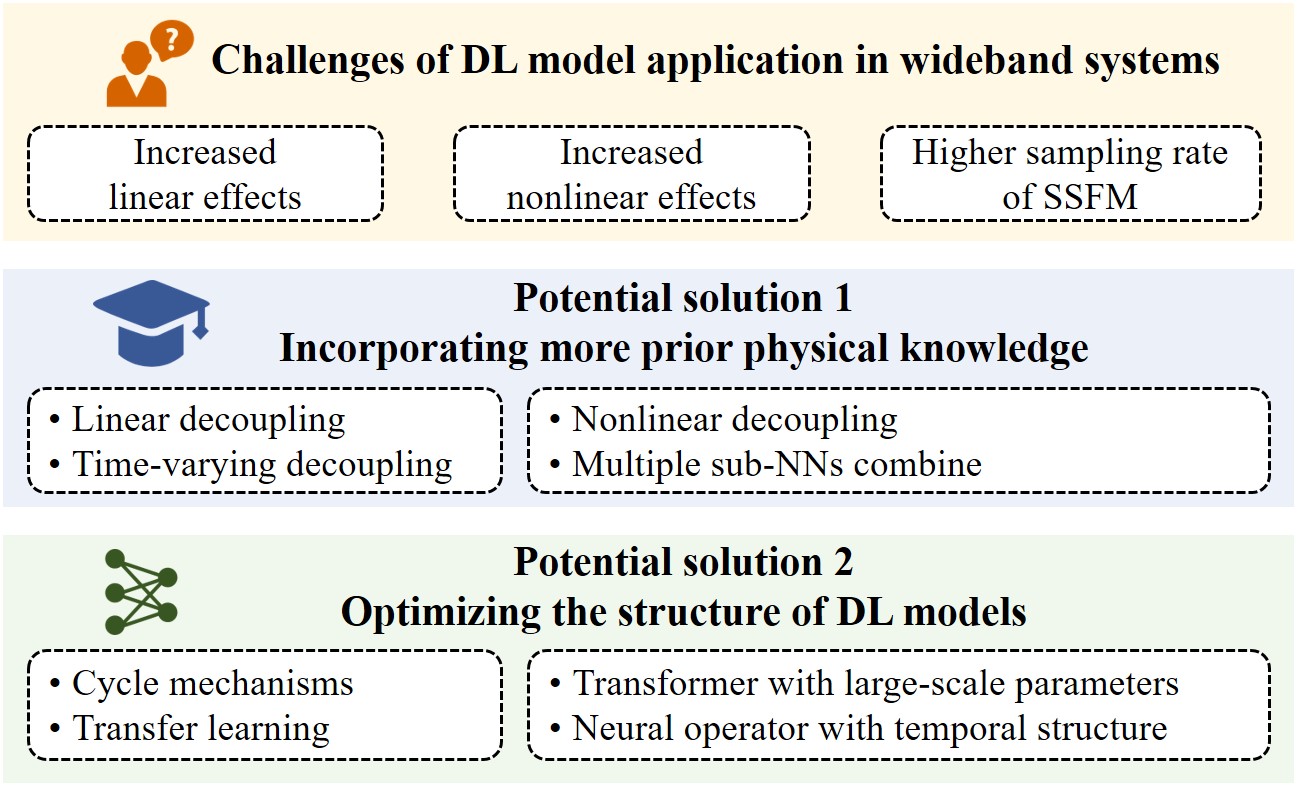}
\caption{Challenges and potential solutions of DL model application in wideband systems.}
\label{fig14}
\end{figure}

The performance of DL-based optical fiber channel modeling deteriorates in configurations with more-channel and higher-rate, posing significant challenges for their application in next-generation wideband optical fiber transmission systems. To address these challenges and enhance their capacity, we propose two potential strategies from the perspectives of incorporating more prior physical knowledge and optimizing the structure of DL models.

\subsection{Incorporating more prior physical knowledge}
\subsubsection{\textbf{Two approaches for incorporating prior physical knowledge}}
\begin{figure}[!t]
\centering
\includegraphics[width=3.3in]{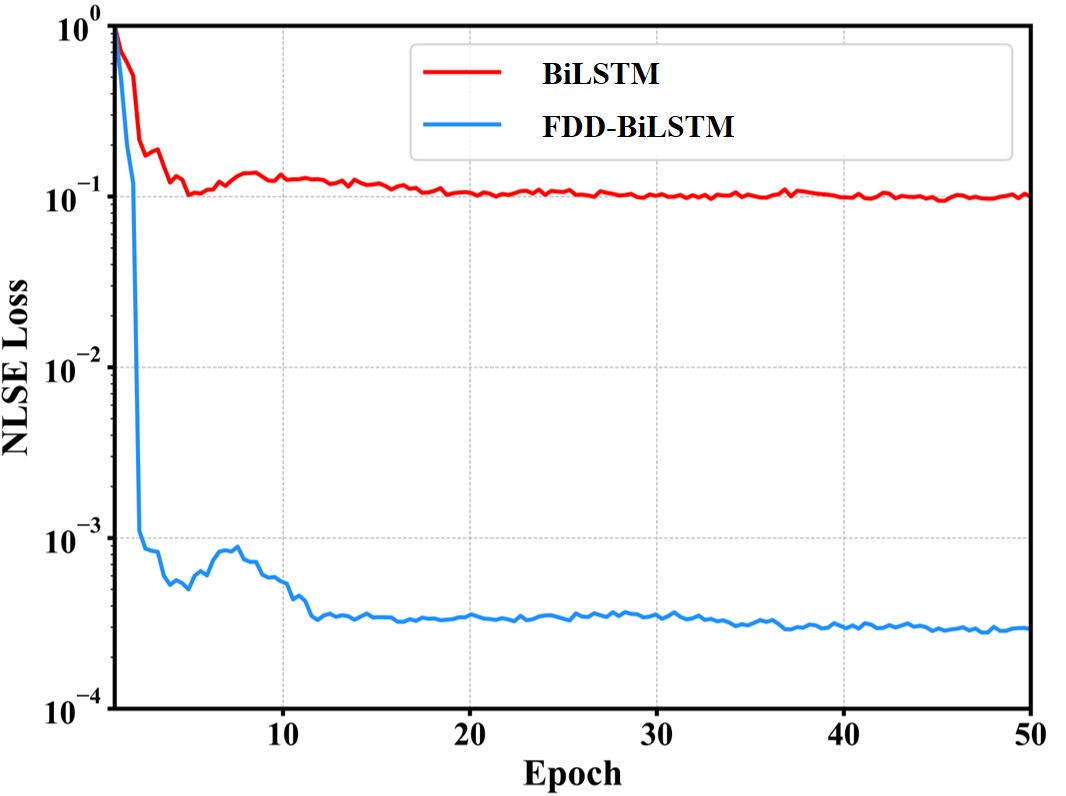}
\caption{The trend of NLSE loss of BiLSTM and FDD-BiLSTM during training.}
\label{fig12}
\end{figure}

The comparison results in this paper demonstrate that FDD-based schemes, employing a distributed architecture combined with a physics-data hybrid-driven approach, outperforms purely data-driven models. These results suggest that incorporating physical prior knowledge have better application prospects. FDD achieves incorporating physical knowledge by decoupling linear effects during the data preprocessing stage. Beyond this approach, another effective strategy is embedding physical constraints into the loss function, as exemplified by PINO \cite{PINO}. By integrating the NLSE directly into the loss function, training can proceed in an unsupervised manner, with the input-output mapping constrained to follow the physical characteristics described by the NLSE. PINO also reduces data requirements and improves training efficiency. Despite their different methods of integration physical knowledge, both FDD and PINO reflect similar underlying physical mechanisms and yield comparable outcomes. In \cite{ACP_ZYF}, the training behavior of FDD-BiLSTM and BiLSTM was analyzed through the evolution of NLSE loss. As illustrated in Fig. \ref{fig12}, FDD-BiLSTM shows a rapid decline in NLSE loss at the early training stages, while BiLSTM converges more slowly. This suggests that models incorporating linear decoupling can inherently learn the NLSE constraint between input and output data, even without explicitly including NLSE loss during training. These observations indicate that both methods---embedding physical prior knowledge through data preprocessing or via the loss function---achieve similar beneficial effects. However, despite these advances, physics-data hybrid-driven methods still suffer from significant accuracy degradation in wideband scenarios. To overcome this limitation, future schemes must incorporate richer physical knowledge by decoupling additional features---not only linear CD but also nonlinear characteristics and time-varying linear impairments — to further enhance model accuracy and robustness in complex wideband systems.

\subsubsection{\textbf{Nonlinear characteristic decoupling}}
As discussed in Section V, the enhanced nonlinear effects in wideband WDM systems significantly limit the performance of DL models, particularly due to the more intricate inter-channel nonlinear effects introduced by an increased number of channels. Decoupling these nonlinear effects offers a promising solution to address this challenge. Current DL schemes typically employ a single model to handle all nonlinearities across multiple channels, which limits the model's ability to effectively capture these complex interactions. A potential solution, based on the characterization of the nonlinear term in the NLSE, is to tackle different nonlinear effects separately. In \cite{DPNO}, stacked DeepONet structures have been employed for optical fiber channel waveform modeling in WDM systems covering the C + L band. This approach employs multiple DeepONets, each designed to fit the intra-channel nonlinearities of individual WDM channels. By isolating the nonlinearities within each channel, this method simplifies the complexity for each DeepONet, enabling more efficient modeling of multi-channel signals. While this method does not directly address inter-channel nonlinearities, it offers a promising direction for future model development. In \cite{DMB_nonlinear_compen}, a strategy involving multiple sub-neural networks was applied to nonlinearity compensation. These sub-networks separately handle intra-channel and inter-channel nonlinearities across various subcarriers, improving accuracy by reducing the learning complexity for each individual network. These results suggest that decoupling nonlinearities and using distinct sub-networks to separately address intra-channel and inter-channel effects can significantly enhance modeling accuracy in wideband WDM systems.

\subsubsection{\textbf{Time-varying linear characteristic decoupling}}
Existing DL models are typically employed as replacements for SSFM, focusing primarily on the static linear and nonlinear effects of optical fibers, while overlooking dynamic linear impairments such as state of polarization fluctuations, polarization mode dispersion, phase noise, and frequency offset introduced by the laser. To more accurately reflect real-world optical transmission systems, these dynamic effects must be incorporated. However, DL models are inherently more adept at capturing deterministic system responses and struggle with modeling dynamic, time-varying behaviors. In contrast, advanced DSP algorithms grounded in physical models---including MIMO equalizers, frequency offset estimation, and carrier phase recovery---excel in compensating for dynamic linear impairments. These algorithms offer a pathway to separate dynamic linear effects from the transmitted signal, enabling the creation of datasets that isolate deterministic nonlinear effects for more effective DL model training. A digital twin framework proposed in \cite{DT} integrates these DSP algorithms with DL models to mirror real optical fiber transmission systems. In this framework, dynamic linear effects are decoupled by DSP modules, producing labels that preserve only residual deterministic effects for neural network training. The demonstrated success of this DT framework provides a promising solution for addressing dynamic linear effects, paving the way for more realistic DL-based optical channel modeling.

\subsection{Optimizing the structure of DL models}
In addition to incorporating more physical prior knowledge, optimizing a DL model structure that conforms to the characteristics of fiber channel is also a potential solution to address the challenges such as longer inter-symbol correlations and stronger nonlinear effects. 


\subsubsection{\textbf{Enhancing the capacity for addressing longer inter-symbol correlations}}
To effectively capture longer inter-symbol correlations, DL models require input sequences enriched with more adjacent symbols, enabling accurate prediction of the target output symbol. However, this leads to significantly longer input windows, resulting in increased model scale, training difficulty, and computational complexity. One promising approach to mitigate these challenges is to reuse computational results from adjacent symbols, thereby reducing redundant calculations. The Co-LSTM \cite{Co_lstm} and Seq2Seq-FDD \cite{Shi:25} architecture introduces a recycling mechanism during inference, reusing hidden states from neighboring LSTM cells to reduce computational repetition and simplify the complexity of conventional LSTM structures. Beyond inference efficiency, improving training efficiency for large-scale models with long input sequences—under limited training data and constrained computational resources—is equally critical. Transfer learning \cite{Transfer_learning} has emerged as an effective strategy, leveraging multi-stage training to address this challenge. Its feasibility in waveform modeling can be justified from two perspectives: the characteristics of optical fiber channels and the structures of temporal DL models. From the perspective of channel characteristics, nonlinear inter-symbol correlations are predominantly governed by the nearest neighboring symbols, while the influence of distant symbols is relatively weaker \cite{many_to_many}. This allows DL models to initially focus on learning the dominant nonlinear effects from nearby symbols and subsequently extend to capture weaker influence from distant symbols via transfer learning. From the perspective of model structure, temporal neural networks such as BiLSTM and Transformer inherently process input sequences in a step-wise or parallel manner across time steps. BiLSTM employs recurrent units with shared parameters to sequentially process each time step, while the Transformer captures inter-symbol correlations through cross-correlations among Query, Key, and Value matrices \cite{Transformer}, with linear transformations of Query, Key, and Value computed in parallel. These designs enable both BiLSTM and Transformer to expand the input sequence length by number of time steps—without introducing additional model parameters. Therefore, leveraging the physical characteristics of nonlinear correlations and the parameter-sharing nature of temporal models, a practical training strategy can be designed: a small-scale model with a short input window is first trained to learn dominant nonlinear effects from the most adjacent symbols. This pre-trained model is then fine-tuned into a large-scale version with a longer input window to capture residual nonlinearities from distant symbols, using transfer learning while maintaining the same parameter. This two-stage approach offers significant advantages: the small-scale model converges faster due to reduced complexity, while the fine-tuned large-scale model inherits the learned features and achieves accelerated convergence and higher final accuracy. Seq2Seq-FDD successfully integrates BiLSTM with transfer learning to implement this multi-stage training paradigm. In a 140-GBaud system, it more efficiently learns inter-symbol correlations and achieves superior modeling accuracy. While this framework has been demonstrated with recurrent architectures, its extension to Transformer-based models—leveraging the self-attention mechanism—remains an open research direction and warrants further investigation.

\subsubsection{\textbf{Enhancing the capacity for fitting stronger nonlinear effects}}
To enhance the nonlinear fitting capacity of DL models and address the challenge of modeling stronger nonlinear effects, increasing the number of model parameters represents a direct and effective strategy. Equally important is the design of architectures capable of efficiently utilizing large parameter counts. Prior studies \cite{Scaling_law} have demonstrated a positive correlation between model scale and accuracy, indicating that larger models generally achieve superior performance. However, for architectures such as LSTM, simply increasing parameter count often yields diminishing returns due to inherent bottlenecks in model capacity \cite{LSTM_VS_Transformer}.In contrast, the Transformer architecture—powered by its multi-head attention mechanism—exhibits exceptional scalability, enabling models with tens of billions of parameters, as exemplified by large language models, such as GPT \cite{GPT1, GPT2, GPT3}. Within this large-scale regime, a consistent performance improvement with increasing parameter count has been observed, including the emergence of capabilities beyond critical scaling thresholds \cite{Scaling_law}. This scalability makes Transformer-based architectures a promising candidate for scaling up DL models to handle increasingly complex nonlinear modeling tasks in optical communications. In our previous analysis, the Transformer model employed a relatively small parameter count, limiting its potential and resulting in a less pronounced accuracy advantage. Future work may achieve improved modeling performance by scaling up the parameter size. Moreover, numerous studies have proposed variants \cite{Transformer_survey} of the self-attention mechanism—such as Vision Transformer \cite{vit} for computer vision and BERT \cite{BERT} for natural language processing—to enhance model accuracy for specific tasks. In the context of optical fiber channel modeling, identifying a self-attention structure better suited to the temporal and nonlinear characteristics of the channel is expected to further improve the Transformer’s modeling fidelity.
Beyond conventional neural networks, neural operators have emerged as a compelling approach due to their strong generalization capabilities. Although, current comparisons indicate that models like DeepONet and FNO have yet to deliver outstanding performance for wideband WDM systems. Their primary limitation lies in their fully connected architectures, which are less suitable for scenarios requiring extended input windows to capture long ISI. A promising direction involves embedding temporal architectures within neural operator frameworks. For example, DeepONet’s flexible trunk and branch network structures could be enhanced by replacing fully connected layers with temporal architectures, enabling them to better capture long inter-symbol correlations and address the challenges posed by stronger nonlinearities and extended ISI in wideband systems.

\section{Conclusion}
In this paper, we review DL-based optical fiber channel waveform modeling schemes and elaborate on key characteristics, inducing overall and distributed schemes, pure data-driven and physics-data hybrid driven schemes, neural networks and neural operators. To enable a fair and comprehensive comparison and to assess the application potential of existing DL solutions in wideband scenarios, we propose a DSP-assisted accuracy evaluation method. This method integrates both waveform error and transmission performance error analysis. Moreover, by incorporating a nonlinear compensation algorithm—-multi-channel DBP—-it resolves the inconsistency between waveform and transmission performance metrics that arises when only linear DSP is applied. This establishes a fair and standardized benchmark for evaluating modeling accuracy. 

Using this evaluation framework, we conduct a comprehensive comparison of DL schemes, spanning from simple few-channel, low-rate configurations to complex wideband scenarios.
The comparison results show that the scheme combining the FDD framework with a temporal neural network achieves an NMSE improvement of 85.6\% and 78.2\% over FNO and DeepONet, respectively, in the 5-channel, 50 GBaud scenario. In more challenging configurations—13-channel at 50 GBaud and 5-channel at 100 GBaud—FDD-BiLSTM outperforms non-FDD-BiLSTM by 83.7\% and 59.3\% in NMSE, respectively. Among all evaluated schemes, the enhanced FDD variant, Seq2Seq-FDD, achieves the best performance, with Q-factor error improvements of 0.64 dB and 0.12 dB over FDD-BiLSTM under the two wideband configurations.

Despite these advantages, we observe that even Seq2Seq-FDD experience accuracy degradation as the number of channels and symbol rate increase. We analyze the underlying challenges from three perspectives: (1) higher sampling rates, (2) enhanced dispersion-induced ISI, and (3) more complex nonlinear interactions. Potential solutions are discussed, including the integration of additional physical priors and structural optimization of DL models. We believe that the standardized evaluation framework, comprehensive comparative results, and in-depth challenge analysis presented in this work will facilitate the rapid advancement of DL-based channel modeling in wideband systems and support the evolution of next-generation optical networks.

\section*{\add{Appendix A}}
\begin{figure}[H]
\centering
\includegraphics[width=3.3in]{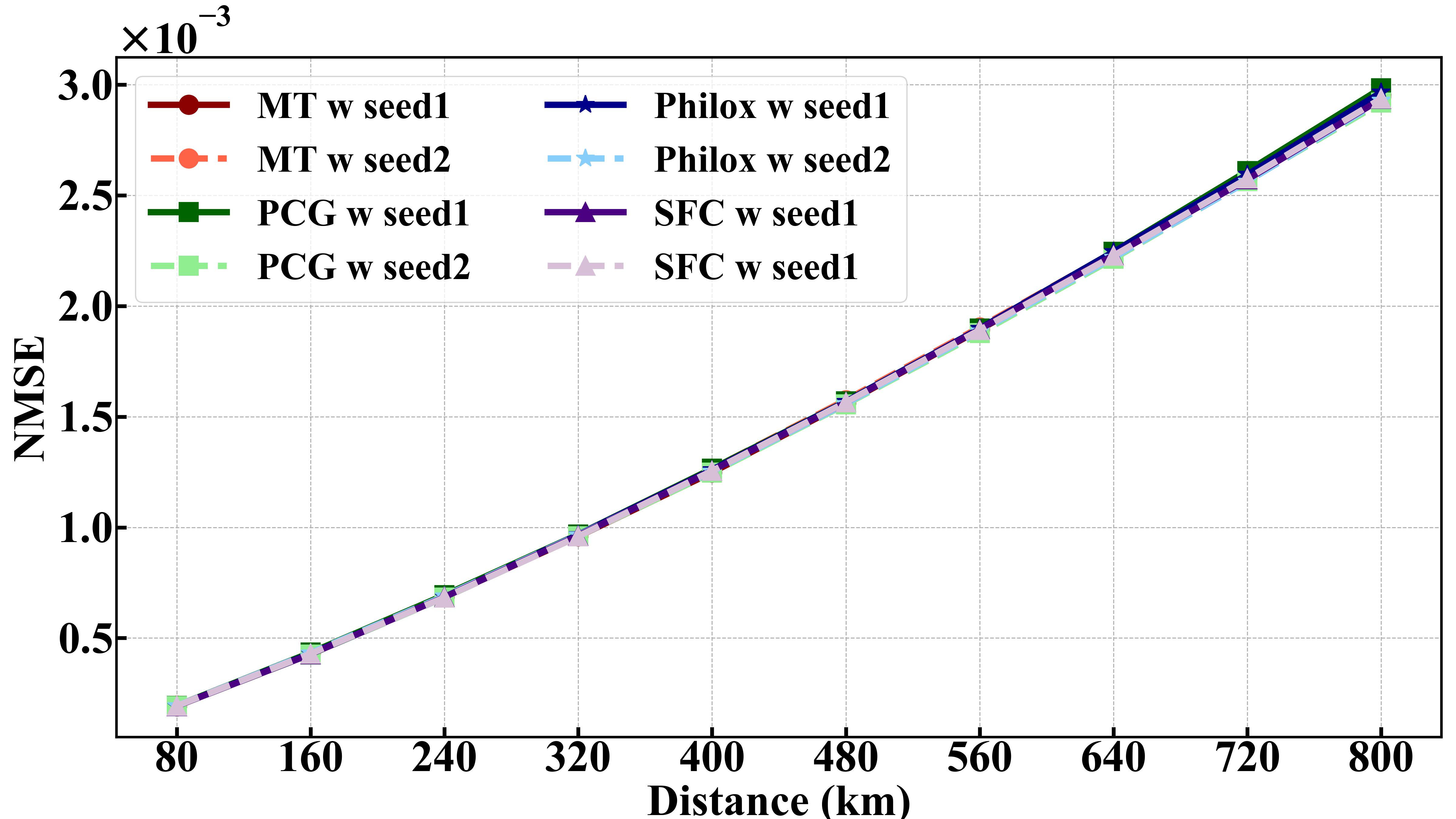}
\caption{\add{The relationship between NMSE and distance with various PRBS generated by various pseudo random number generators.}}
\label{PRBS}
\end{figure}

\add{We evaluate the impact of different PRBS on the test results. The model used for this evaluation is the FDD-BiLSTM scheme illustrated in Fig. \ref{fig7}. We employ eight distinct PRBS generated by four different pseudo random number generators: Mersenne Twister (MT) algorithm, the Permuted Congruential Generator (PCG), the Philox algorithm, and the Small Fast Counter (SFC) algorithm. Two pseudo random number seeds are respectively used to drive these generators.} Fig. \ref{PRBS} \add{illustrates the NMSE variation with distance for the different PRBS. It is evident that the DL model exhibits consistent performance trends across all sequences. At 800 km, the standard deviation of the NMSE across the eight PRBSs is 1.95e-5, confirming that the choice of PRBS has negligible impact on the final results.}

\section*{\add{Appendix B}}
\add{The generalization capability of DL models is also of significant importance. The parameters requiring generalization can be categorized into two groups: signal parameters and channel parameters. Signal parameters typically include modulation format, launch power, and symbol rate, while channel parameters encompass transmission distance, span length, fiber dispersion, and attenuation coefficient. Numerous studies} \cite{BiLSTM, FDD, Co_lstm, DPNO} \add{have demonstrated that DL models exhibit strong generalization performance for modulation formats and launch power. Since launch power directly influences the nonlinear characteristics of the optical fiber channel, it is generally necessary to construct training datasets spanning a range of power levels to achieve robust generalization. In contrast, generalization across varying symbol rates remains an open challenge and requires further research. With respect to transmission distance, distributed modeling schemes enable flexible generalization by adjusting the number of cascaded DL models. For finer-grained generalization of arbitrary span length, Zeng et al.} \cite{embedding} \add{introduced a parameter encoding architecture that enhances the model’s adaptability to variable span lengths. For fiber attenuation, similar to launch power, changes in the attenuation coefficient alter the signal power evolution along the fiber, thereby affecting the strength of nonlinear impairments. Consequently, achieving generalization over different attenuation conditions requires training data that includes diverse attenuation coefficients. In the case of fiber dispersion, the FDD scheme models linear dispersion effects using a physical model. By flexibly adjusting the dispersion parameter within the physical model, the FDD framework can achieve excellent generalization of varying dispersion coefficients.}
\section*{\add{Appendix C}}

\begin{figure}[H]
\centering
\includegraphics[width=3.3in]{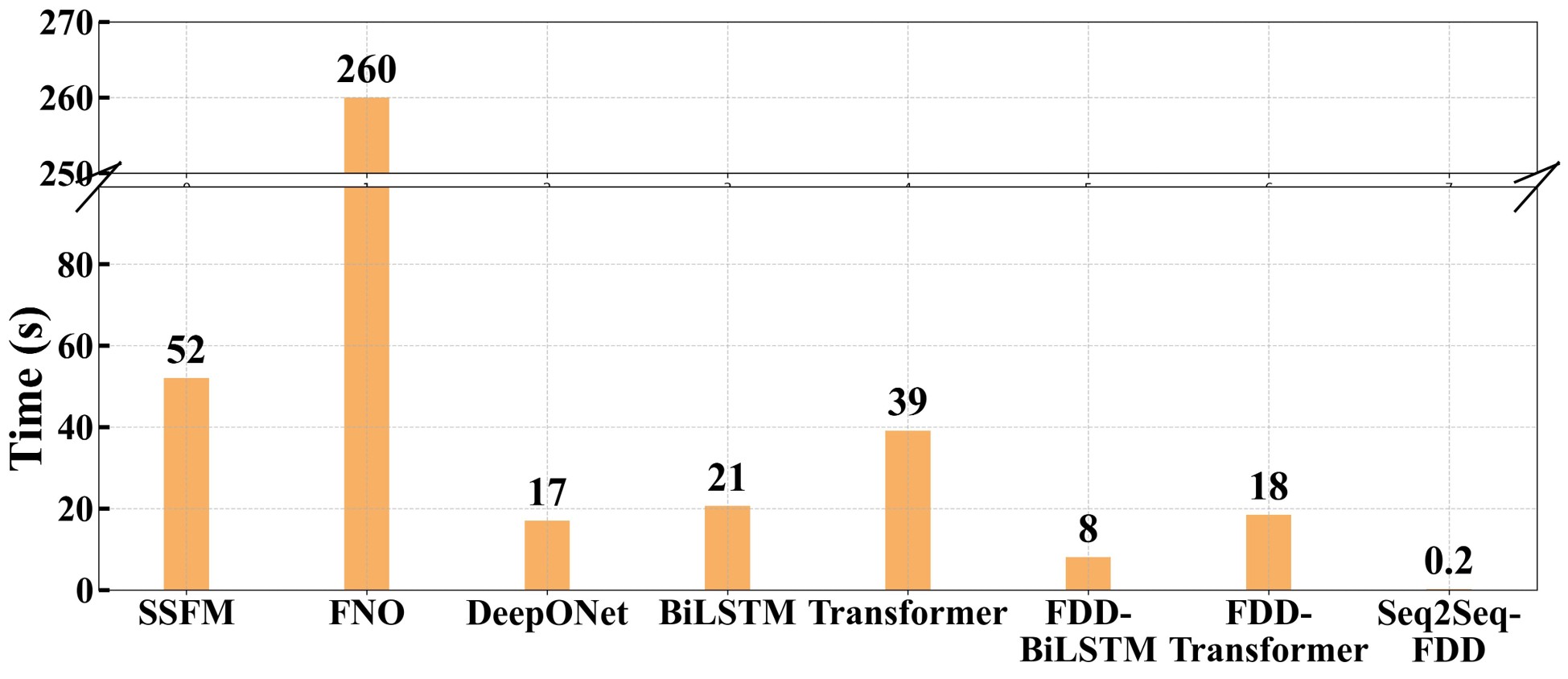}
\caption{\add{DL model inference time after single-span transmission in the 5-channel, 50 GBaud scenario.}}
\label{Time}
\end{figure}

\add{We evaluate the inference time of each DL model in the 5-channel, 50 GBaud scenario. Fig.} \ref{Time} \add{exhibits the inference time of DL models after single-span transmission. Overall, most DL models, except for FNO, exhibit significantly lower inference time compared to SSFM. The FDD-based scheme achieves lower computational complexity than its non-FDD counterpart due to the use of a reduced input window. Notably, the Seq2Seq-FDD scheme, which employs a multi-input multi-output architecture, reduces redundant computations from padding symbols. As a result, it achieves the lowest computational complexity, particularly in multi-channel and high-speed scenarios.}

\section*{Acknowledgments}
The authors acknowledge the funding provided by the National Key R\&D Program of China (2023YFB2905400), National Natural Science Foundation of China (62025503), and Shanghai Jiao Tong University 2030 Initiative.

\bibliographystyle{IEEEtran}
\bibliography{IEEEabrv,ref}

\begin{thebibliography}{10}
\providecommand{\url}[1]{#1}
\csname url@samestyle\endcsname
\providecommand{\newblock}{\relax}
\providecommand{\bibinfo}[2]{#2}
\providecommand{\BIBentrySTDinterwordspacing}{\spaceskip=0pt\relax}
\providecommand{\BIBentryALTinterwordstretchfactor}{4}
\providecommand{\BIBentryALTinterwordspacing}{\spaceskip=\fontdimen2\font plus
\BIBentryALTinterwordstretchfactor\fontdimen3\font minus \fontdimen4\font\relax}
\providecommand{\BIBforeignlanguage}[2]{{%
\expandafter\ifx\csname l@#1\endcsname\relax
\typeout{** WARNING: IEEEtran.bst: No hyphenation pattern has been}%
\typeout{** loaded for the language `#1'. Using the pattern for}%
\typeout{** the default language instead.}%
\else
\language=\csname l@#1\endcsname
\fi
#2}}
\providecommand{\BIBdecl}{\relax}
\BIBdecl

\bibitem{Multi-Band1}
A.~Ferrari \emph{et~al.}, ``Assessment on the achievable throughput of multi-band {ITU-T G.652.D} fiber transmission systems,'' \emph{J. Lightw. Technol.}, vol.~38, no.~16, pp. 4279--4291, 2020.

\bibitem{Multi-Band2}
M.~Zuo \emph{et~al.}, ``Field trial of real-time $80\text{-}\lambda\times400\text{-}\text{Gb}/\mathrm{s}$ single-carrier {128-GBd DP-QPSK} transmission covering {12-THz C+L} band over 2502-km terrestrial {G.652.D} fibre,'' in \emph{49th European Conference on Optical Communications (ECOC 2023)}, vol. 2023, 2023, pp. 843--846.

\bibitem{Multi-Band3}
D.~Ge \emph{et~al.}, ``Fully-loaded $80\times 400\text{Gb}/\mathrm{s}$ {DP-QPSK} transmission with commercial {12-THz C6T+L6T} {EDFAs} over record distance of 7000km,'' in \emph{2023 Asia Communications and Photonics Conference/2023 International Photonics and Optoelectronics Meetings (ACP/POEM)}, 2023, pp. 1--4.

\bibitem{Multi-Band4}
S.~Escobar-Landero, A.~Lorences-Riesgo, X.~Zhao, Y.~Frignac, and G.~Charlet, ``{S+C+L} high-capacity transmission systems: Challenges and opportunities,'' \emph{J. Lightw. Technol.}, vol.~42, no.~12, pp. 4260--4270, 2024.

\bibitem{Multi-Band5}
Y.~Frignac \emph{et~al.}, ``Record 158.4 {Tb/s} transmission over 2x60 km field {SMF} using {S+C+L 18Thz}-bandwidth lumped amplification,'' in \emph{49th European Conference on Optical Communications (ECOC 2023)}, vol. 2023, 2023, pp. 550--553.

\bibitem{network1}
\BIBentryALTinterwordspacing
Y.~Zhang \emph{et~al.}, ``Optical power control for {GSNR} optimization based on {C+L}-band digital twin systems,'' \emph{J. Lightw. Technol.}, vol.~42, no.~1, pp. 95--105, Jan 2024.
\BIBentrySTDinterwordspacing

\bibitem{network2}
B.~Correia, R.~Sadeghi, E.~Virgillito, A.~Napoli, N.~Costa, J.~Pedro, and V.~Curri, ``Optical power control strategies for optimized {C+L+S}-bands network performance,'' in \emph{2021 Optical Fiber Communications Conference and Exhibition (OFC)}, 2021, pp. 1--3.

\bibitem{network3}
I.~Roberts, J.~M. Kahn, J.~Harley, and D.~W. Boertjes, ``Channel power optimization of {WDM} systems following gaussian noise nonlinearity model in presence of stimulated raman scattering,'' \emph{J. Lightw. Technol.}, vol.~35, no.~23, pp. 5237--5249, 2017.

\bibitem{network4}
Y.~Song, Q.~Fan, C.~Lu, D.~Wang, and A.~P.~T. Lau, ``Efficient three-step amplifier configuration algorithm for dynamic {C+L}-band links in presence of stimulated raman scattering,'' \emph{J. Lightw. Technol.}, vol.~41, no.~5, pp. 1445--1453, 2023.

\bibitem{network5}
C.~Zhang \emph{et~al.}, ``Potential failure cause identification for optical networks using deep learning with an attention mechanism,'' \emph{{IEEE} J. Opt. Commun. Netw.}, vol.~14, no.~2, pp. A122--A133, 2022.

\bibitem{network6}
Y.~Zhang \emph{et~al.}, ``Building a digital twin for large-scale and dynamic {C+L}-band optical networks,'' \emph{{IEEE} J. Opt. Commun. Netw.}, vol.~15, no.~12, pp. 985--998, 2023.

\bibitem{nonlinear1}
A.~Napoli \emph{et~al.}, ``Reduced complexity digital back-propagation methods for optical communication systems,'' \emph{J. Lightw. Technol.}, vol.~32, no.~7, pp. 1351--1362, 2014.

\bibitem{nonlinear2}
O.~Vassilieva, I.~Kim, and T.~Ikeuchi, ``Enabling technologies for fiber nonlinearity mitigation in high capacity transmission systems,'' \emph{J. Lightw. Technol.}, vol.~37, no.~1, pp. 50--60, 2019.

\bibitem{nonlinear3}
Q.~Fan, G.~Zhou, T.~Gui, C.~Lu, and A.~P.~T. Lau, ``Advancing theoretical understanding and practical performance of signal processing for nonlinear optical communications through machine learning,'' \emph{Nat. Commun.}, vol.~11, p. 3694, 07 2020.

\bibitem{nonlinear4}
O.~Sidelnikov, A.~Redyuk, S.~Sygletos, M.~Fedoruk, and S.~Turitsyn, ``Advanced convolutional neural networks for nonlinearity mitigation in long-haul {WDM} transmission systems,'' \emph{J. Lightw. Technol.}, vol.~39, no.~8, pp. 2397--2406, 2021.

\bibitem{nonlinear5}
X.~Lin, S.~Luo, S.~K.~O. Soman, O.~A. Dobre, L.~Lampe, D.~Chang, and C.~Li, ``Perturbation theory-aided learned digital back-propagation scheme for optical fiber nonlinearity compensation,'' \emph{J. Lightw. Technol.}, vol.~40, no.~7, pp. 1981--1988, 2022.

\bibitem{LDSP}
Z.~Niu, H.~Yang, L.~Li, M.~Shi, G.~Xu, W.~Hu, and L.~Yi, ``Learnable digital signal processing: a new benchmark of linearity compensation for optical fiber communications,'' \emph{Light Sci. Appl.}, vol.~13, no.~1, p. 188, 2024.

\bibitem{e2e1}
B.~Karanov \emph{et~al.}, ``End-to-end deep learning of optical fiber communications,'' \emph{J. Lightw. Technol.}, vol.~36, no.~20, pp. 4843--4855, 2018.

\bibitem{e2e2}
\BIBentryALTinterwordspacing
B.~Karanov, D.~Lavery, P.~Bayvel, and L.~Schmalen, ``End-to-end optimized transmission over dispersive intensity-modulated channels using bidirectional recurrent neural networks,'' \emph{Opt. Express}, vol.~27, no.~14, pp. 19\,650--19\,663, Jul 2019.
\BIBentrySTDinterwordspacing

\bibitem{e2e3}
B.~Karanov, M.~Chagnon, V.~Aref, D.~Lavery, P.~Bayvel, and L.~Schmalen, ``Optical fiber communication systems based on end-to-end deep learning : (invited paper),'' in \emph{2020 IEEE Photonics Conference (IPC)}, 2020, pp. 1--2.

\bibitem{e2e4}
Z.~Niu, H.~Yang, H.~Zhao, C.~Dai, W.~Hu, and L.~Yi, ``End-to-end deep learning for long-haul fiber transmission using differentiable surrogate channel,'' \emph{J. Lightw. Technol.}, vol.~40, no.~9, pp. 2807--2822, 2022.

\bibitem{e2e5}
Y.~Xu, L.~Huang, W.~Jiang, X.~Guan, W.~Hu, and L.~Yi, ``End-to-end learning for {100G-PON} based on noise adaptation network,'' \emph{J. Lightw. Technol.}, vol.~42, no.~7, pp. 2328--2337, 2024.

\bibitem{e2e6}
M.~Li and S.~Wang, ``End-to-end learning for chromatic dispersion compensation in optical fiber communication,'' \emph{IEEE Commun. Lett.}, vol.~26, no.~8, pp. 1829--1832, 2022.

\bibitem{e2e7}
J.~Song, C.~Häger, J.~Schröder, A.~G.~I. Amat, and H.~Wymeersch, ``Model-based {E}nd-to-{E}nd learning for {WDM} systems with transceiver hardware impairments,'' \emph{IEEE J. Sel. Top. Quantum Electron}, vol.~28, no. 4: Mach. Learn. in Photon. Commun. and Meas. Syst., pp. 1--14, 2022.

\bibitem{Book_OFC}
\BIBentryALTinterwordspacing
D.~Gloge, ``Optical fibers for communication,'' \emph{Appl. Opt.}, vol.~13, no.~2, pp. 249--254, Feb 1974.
\BIBentrySTDinterwordspacing

\bibitem{GN_Model}
P.~Poggiolini, ``The {GN} model of non-linear propagation in uncompensated coherent optical systems,'' \emph{J. Lightw. Technol.}, vol.~30, no.~24, pp. 3857--3879, 2012.

\bibitem{EGN}
\BIBentryALTinterwordspacing
A.~Carena, G.~Bosco, V.~Curri, Y.~Jiang, P.~Poggiolini, and F.~Forghieri, ``{EGN} model of non-linear fiber propagation,'' \emph{Opt. Express}, vol.~22, no.~13, pp. 16\,335--16\,362, Jun 2014.
\BIBentrySTDinterwordspacing

\bibitem{Book_NFO}
G.~P. Agrawal, ``Nonlinear fiber optics,'' in \emph{Nonlinear Science at the Dawn of the 21st Century}, P.~L. Christiansen, M.~P. S{\o}rensen, and A.~C. Scott, Eds.\hskip 1em plus 0.5em minus 0.4em\relax Berlin, Heidelberg: Springer Berlin Heidelberg, 2000, pp. 195--211.

\bibitem{SSFM_four_power_growth}
P.~Serena, C.~Lasagni, S.~Musetti, and A.~Bononi, ``On numerical simulations of ultra-wideband long-haul optical communication systems,'' \emph{J. Lightw. Technol.}, vol.~38, no.~5, pp. 1019--1031, 2020.

\bibitem{universal_approximators}
\BIBentryALTinterwordspacing
K.~Hornik, M.~Stinchcombe, and H.~White, ``Multilayer feedforward networks are universal approximators,'' \emph{Neural Networks}, vol.~2, no.~5, pp. 359--366, 1989.
\BIBentrySTDinterwordspacing

\bibitem{parallel}
J.~Wei, X.~Zhang, Z.~Ji \emph{et~al.}, ``Deploying and scaling distributed parallel deep neural networks on the tianhe-3 prototype system,'' \emph{Sci. Rep.}, vol.~11, p. 20244, 2021.

\bibitem{BiLSTM}
D.~Wang \emph{et~al.}, ``Data-driven optical fiber channel modeling: A deep learning approach,'' \emph{J. Lightw. Technol.}, vol.~38, no.~17, pp. 4730--4743, 2020.

\bibitem{CGAN}
H.~Yang, Z.~Niu, S.~Xiao, J.~Fang, Z.~Liu, D.~Fainsin, and L.~Yi, ``Fast and accurate optical fiber channel modeling using generative adversarial network,'' \emph{J. Lightw. Technol.}, vol.~39, no.~5, pp. 1322--1333, 2021.

\bibitem{multi-attention}
Y.~Zang, Z.~Yu, K.~Xu, M.~Chen, S.~Yang, and H.~Chen, ``Multi-span long-haul fiber transmission model based on cascaded neural networks with multi-head attention mechanism,'' \emph{J. Lightw. Technol.}, vol.~40, no.~19, pp. 6347--6358, 2022.

\bibitem{Zang:22}
\BIBentryALTinterwordspacing
------, ``Data-driven fiber model based on the deep neural network with multi-head attention mechanism,'' \emph{Opt. Express}, vol.~30, no.~26, pp. 46\,626--46\,648, Dec 2022.
\BIBentrySTDinterwordspacing

\bibitem{FNO_1}
X.~He \emph{et~al.}, ``Fourier neural operator for accurate optical fiber modeling with low complexity,'' \emph{J. Lightw. Technol.}, vol.~41, no.~8, pp. 2301--2311, 2023.

\bibitem{Co_lstm}
\BIBentryALTinterwordspacing
J.~Zheng, T.~Zhang, and F.~Zhang, ``Co-lstm-based fiber link modeling with ase noise tracking for long-haul coherent optical transmission,'' \emph{Opt. Lett.}, vol.~49, no.~7, pp. 1848--1851, Apr 2024.
\BIBentrySTDinterwordspacing

\bibitem{PINO}
Y.~Song, D.~Wang, Q.~Fan, X.~Jiang, X.~Luo, and M.~Zhang, ``Physics-informed neural operator for fast and scalable optical fiber channel modelling in multi-span transmission,'' in \emph{2022 European Conference on Optical Communication (ECOC)}, 2022, pp. 1--4.

\bibitem{DNN}
R.~Jiang, Z.~Fu, Y.~Bao, H.~Wang, X.~Ding, and Z.~Wang, ``Data-driven method for nonlinear optical fiber channel modeling based on deep neural network,'' \emph{{IEEE} Photon. J.}, vol.~14, no.~4, pp. 1--8, 2022.

\bibitem{FNO_5}
Q.~Qiu, H.~Lun, X.~Liu, L.~Yi, W.~Hu, and Q.~Zhuge, ``Fourier neural operator based fibre channel modelling for optical transmission,'' in \emph{2022 European Conference on Optical Communication (ECOC)}, 2022, pp. 1--4.

\bibitem{FDD}
H.~Yang, Z.~Niu, H.~Zhao, S.~Xiao, W.~Hu, and L.~Yi, ``Fast and accurate waveform modeling of long-haul multi-channel optical fiber transmission using a hybrid model-data driven scheme,'' \emph{J. Lightw. Technol.}, vol.~40, no.~14, pp. 4571--4580, 2022.

\bibitem{ACP}
M.~Shi, H.~Yang, Z.~Niu, C.~Zeng, S.~Xiao, W.~Hu, and L.~Yi, ``Accurate and efficient optical fiber {WDM} transmission modeling using the encoder-only transformer with feature decoupling distributed method,'' in \emph{2023 Asia Communications and Photonics Conference/2023 International Photonics and Optoelectronics Meetings (ACP/POEM)}, 2023, pp. 1--5.

\bibitem{DPNO}
X.~Zhang, M.~Zhang, Y.~Song, X.~Jiang, F.~Zhang, and D.~Wang, ``Deeponet-based waveform-level simulation for a wideband nonlinear {WDM} system,'' \emph{J. Lightw. Technol.}, vol.~41, no.~22, pp. 6908--6922, 2023.

\bibitem{Shi:25}
\BIBentryALTinterwordspacing
M.~Shi \emph{et~al.}, ``Fast and accurate waveform modeling based on sequence-to-sequence framework for multi-channel and high-rate optical fiber transmission,'' \emph{Opt. Lett.}, vol.~50, no.~7, pp. 2286--2289, Apr 2025.
\BIBentrySTDinterwordspacing

\bibitem{OSNR_impact}
G.~Ye, J.~Xiang, G.~Zhou, M.~Xiang, J.~Li, Y.~Qin, and S.~Fu, ``Impact of the input {OSNR} on data-driven optical fiber channel modeling,'' \emph{{IEEE} J. Opt. Commun. Netw.}, vol.~15, no.~2, pp. 78--86, 2023.

\bibitem{embedding}
C.~Zeng, Z.~Niu, H.~Yang, M.~Shi, W.~Hu, and L.~Yi, ``Enhancing generalization in neural channel model for optical fiber {WDM} transmission through learned encoding of system parameters,'' in \emph{2024 Optical Fiber Communications Conference and Exhibition (OFC)}, 2024, pp. 1--3.

\bibitem{fiber_type}
\BIBentryALTinterwordspacing
R.~Jiang, Z.~Wang, T.~Jia, Z.~Fu, C.~Shang, and C.~Wu, ``Flexible optical fiber channel modeling based on a neural network module,'' \emph{Opt. Lett.}, vol.~48, no.~16, pp. 4332--4335, Aug 2023.
\BIBentrySTDinterwordspacing

\bibitem{Multi_core}
M.~Ma, H.~Chang, R.~Gao, D.~Guo, X.~Liu, and M.~Yuan, ``Modeling of multi-core fiber channel based on m-cgan for high capacity fiber optical communication,'' in \emph{2023 Asia Communications and Photonics Conference/2023 International Photonics and Optoelectronics Meetings (ACP/POEM)}, 2023, pp. 01--05.

\bibitem{few_mode}
M.~Yuan \emph{et~al.}, ``A conditional generative adversarial network aided few-mode fiber channel modeling for large-capacity optical fiber communication,'' in \emph{2023 21st International Conference on Optical Communications and Networks (ICOCN)}, 2023, pp. 1--3.

\bibitem{RoF}
Y.~Zhu, J.~Ye, L.~Yan, T.~Zhou, P.~Li, X.~Zou, and W.~Pan, ``Transformer-based high-fidelity modeling method for radio over fiber link,'' \emph{J. Lightw. Technol.}, vol.~41, no.~9, pp. 2657--2665, 2023.

\bibitem{OFDM}
N.~Zhang, H.~Yang, Z.~Niu, L.~Zheng, C.~Chen, S.~Xiao, and L.~Yi, ``Transformer-based long distance fiber channel modeling for optical {OFDM} systems,'' \emph{J. Lightw. Technol.}, vol.~40, no.~24, pp. 7779--7789, 2022.

\bibitem{FSO}
W.~Chen \emph{et~al.}, ``Deep learning-based channel modeling for free space optical communications,'' \emph{J. Lightw. Technol.}, vol.~41, no.~1, pp. 183--198, 2023.

\bibitem{NN1}
S.~Boscolo and C.~Finot, ``Artificial neural networks for nonlinear pulse shaping in optical fibers,'' \emph{Opt. Laser Technol.}, vol. 131, p. 106439, 2020.

\bibitem{NN2}
T.~Zahavy, A.~Dikopoltsev, D.~Moss, G.~I. Haham, O.~Cohen, S.~Mannor, and M.~Segev, ``Deep learning reconstruction of ultrashort pulses,'' \emph{Optica}, vol.~5, no.~5, pp. 666--673, 2018.

\bibitem{NN3}
L.~Salmela, N.~Tsipinakis, A.~Foi, C.~Billet, J.~M. Dudley, and G.~Genty, ``Predicting ultrafast nonlinear dynamics in fibre optics with a recurrent neural network,'' \emph{Nat. Mach. Intell.}, vol.~3, no.~4, pp. 344--354, 2021.

\bibitem{NN4}
H.~Yang, H.~Zhao, Z.~Niu, G.~Pu, S.~Xiao, W.~Hu, and L.~Yi, ``Low-complexity full-field ultrafast nonlinear dynamics prediction by a convolutional feature separation modeling method,'' \emph{Opt. Express}, vol.~30, no.~24, pp. 43\,691--43\,705, 2022.

\bibitem{NN5}
U.~Te{\u{g}}in, B.~Rahmani, E.~Kakkava, N.~Borhani, C.~Moser, and D.~Psaltis, ``Controlling spatiotemporal nonlinearities in multimode fibers with deep neural networks,'' \emph{Apl Photonics}, vol.~5, no.~3, 2020.

\bibitem{NN6}
M.~Raissi, P.~Perdikaris, and G.~E. Karniadakis, ``Physics-informed neural networks: A deep learning framework for solving forward and inverse problems involving nonlinear partial differential equations,'' \emph{J. Comput. Phys.}, vol. 378, pp. 686--707, 2019.

\bibitem{NN7}
X.-M. Liu, Z.-Y. Zhang, and W.-J. Liu, ``Physics-informed neural network method for predicting soliton dynamics supported by complex parity-time symmetric potentials,'' \emph{Chin. Phys. Lett.}, vol.~40, no.~7, p. 070501, 2023.

\bibitem{NN8}
S.-Y. Xu, Q.~Zhou, and W.~Liu, ``Prediction of soliton evolution and equation parameters for nls--mb equation based on the phpinn algorithm,'' \emph{Nonlinear Dyn.}, vol. 111, no.~19, pp. 18\,401--18\,417, 2023.

\bibitem{NN9}
X.~Jiang, D.~Wang, Q.~Fan, M.~Zhang, C.~Lu, and A.~P.~T. Lau, ``Solving the nonlinear schr{\"o}dinger equation in optical fibers using physics-informed neural network,'' in \emph{Optical fiber communication conference}.\hskip 1em plus 0.5em minus 0.4em\relax Optica Publishing Group, 2021, pp. M3H--8.

\bibitem{NN10}
D.~Wang, X.~Jiang, Y.~Song, M.~Fu, Z.~Zhang, X.~Chen, and M.~Zhang, ``Applications of physics-informed neural network for optical fiber communications,'' \emph{IEEE Commun. Mag.}, vol.~60, no.~9, pp. 32--37, 2022.

\bibitem{NN11}
X.~Jiang, D.~Wang, Q.~Fan, M.~Zhang, C.~Lu, and A.~P.~T. Lau, ``Physics-informed neural network for nonlinear dynamics in fiber optics,'' \emph{Laser Photonics Rev.}, vol.~16, no.~9, p. 2100483, 2022.

\bibitem{CNLSE}
C.~Menyuk, ``Nonlinear pulse propagation in birefringent optical fibers,'' \emph{IEEE J. Quantum Electron.}, vol.~23, no.~2, pp. 174--176, 1987.

\bibitem{Manakov1}
D.~Marcuse, C.~Manyuk, and P.~Wai, ``Application of the manakov-pmd equation to studies of signal propagation in optical fibers with randomly varying birefringence,'' \emph{J. Lightw. Technol.}, vol.~15, no.~9, pp. 1735--1746, 1997.

\bibitem{Manakov2}
S.~Evangelides, L.~Mollenauer, J.~Gordon, and N.~Bergano, ``Polarization multiplexing with solitons,'' \emph{J. Lightw. Technol.}, vol.~10, no.~1, pp. 28--35, 1992.

\bibitem{symmetric_SSFM}
J.~Shao, X.~Liang, and S.~Kumar, ``Comparison of split-step {F}ourier schemes for simulating fiber optic communication systems,'' \emph{{IEEE} Photon. J.}, vol.~6, no.~4, pp. 1--15, 2014.

\bibitem{ssfm_nonlinear_phase}
O.~Sinkin, R.~Holzlohner, J.~Zweck, and C.~Menyuk, ``Optimization of the split-step {F}ourier method in modeling optical-fiber communications systems,'' \emph{J. Lightw. Technol.}, vol.~21, no.~1, pp. 61--68, 2003.

\bibitem{DBP_origin}
E.~Ip and J.~M. Kahn, ``Compensation of dispersion and nonlinear impairments using digital backpropagation,'' \emph{J. Lightw. Technol.}, vol.~26, no.~20, pp. 3416--3425, 2008.

\bibitem{GAN_challenge}
D.~Saxena, J.~Cao \emph{et~al.}, ``Generative adversarial networks ({GAN}s): challenges, solutions, and future directions,'' \emph{ACM Comput. Surv.}, vol.~54, no.~3, pp. 1--42, 2022.

\bibitem{gan_review}
\BIBentryALTinterwordspacing
J.~Gui, Z.~Sun, Y.~Wen, D.~Tao, and J.~Ye, ``A review on generative adversarial networks: Algorithms, theory, and applications,'' 2020. [Online]. Available: \url{https://arxiv.org/abs/2001.06937}
\BIBentrySTDinterwordspacing

\bibitem{operator_theory1}
T.~Chen and H.~Chen, ``Approximation capability to functions of several variables, nonlinear functionals, and operators by radial basis function neural networks,'' \emph{IEEE Trans. Neural Networks}, vol.~6, no.~4, pp. 904--910, 1995.

\bibitem{operator_theory2}
S.~Wang, H.~Wang, and P.~Perdikaris, ``Learning the solution operator of parametric partial differential equations with physics-informed deeponets,'' \emph{Sci. Adv.}, vol.~7, no.~40, p. eabi8605, 2021.

\bibitem{operator_theory3}
L.~Lu, P.~Jin, G.~Pang, Z.~Zhang, and G.~E. Karniadakis, ``Learning nonlinear operators via deeponet based on the universal approximation theorem of operators,'' \emph{Nat. Mach. Intell.}, vol.~3, no.~3, pp. 218--229, 2021.

\bibitem{overfitting1}
T.~A. Eriksson, H.~Bülow, and A.~Leven, ``Applying neural networks in optical communication systems: Possible pitfalls,'' \emph{IEEE Photonics Technol. Lett.}, vol.~29, no.~23, pp. 2091--2094, 2017.

\bibitem{overfitting2}
L.~Yi, T.~Liao, L.~Huang, L.~Xue, P.~Li, and W.~Hu, ``Machine learning for 100 ${Gb/s/\lambda}$ passive optical network,'' \emph{J. Lightw. Technol.}, vol.~37, no.~6, pp. 1621--1630, 2019.

\bibitem{Adam}
\BIBentryALTinterwordspacing
D.~P. Kingma and J.~Ba, ``Adam: A method for stochastic optimization,'' 2017. [Online]. Available: \url{https://arxiv.org/abs/1412.6980}
\BIBentrySTDinterwordspacing

\bibitem{SGDR}
\BIBentryALTinterwordspacing
I.~Loshchilov and F.~Hutter, ``Sgdr: Stochastic gradient descent with warm restarts,'' 2017. [Online]. Available: \url{https://arxiv.org/abs/1608.03983}
\BIBentrySTDinterwordspacing

\bibitem{FEC_threshold}
\BIBentryALTinterwordspacing
J.~Zhang, X.~Li, Y.~Hu, M.~S. Alam, and D.~V. Plant, ``Spectrally efficient integrated silicon photonic phase-diverse {DD} receiver with near-ideal phase response for c-band {DWDM} transmission,'' \emph{Laser Photonics Rev.}, vol.~18, no.~3, p. 2300623, 2024.
\BIBentrySTDinterwordspacing

\bibitem{perturbation}
Z.~Tao, L.~Dou, W.~Yan, L.~Li, T.~Hoshida, and J.~C. Rasmussen, ``Multiplier-free intrachannel nonlinearity compensating algorithm operating at symbol rate,'' \emph{J. Lightw. Technol.}, vol.~29, no.~17, pp. 2570--2576, 2011.

\bibitem{ACP_ZYF}
Y.~Zhang, Z.~Niu, M.~Shi, W.~Hu, and L.~Yi, ``Improve the fitting accuracy of deep learning for the nonlinear schrödinger equation using linear feature decoupling method,'' in \emph{2024 Asia Communications and Photonics Conference (ACP) and International Conference on Information Photonics and Optical Communications (IPOC)}, 2024, pp. 1--4.

\bibitem{DMB_nonlinear_compen}
\BIBentryALTinterwordspacing
A.~Bakhshali, H.~Najafi, B.~B. Hamgini, and Z.~Zhang, ``Neural network architectures for optical channel nonlinear compensation in digital subcarrier multiplexing systems,'' \emph{Opt. Express}, vol.~31, no.~16, pp. 26\,418--26\,434, Jul 2023.
\BIBentrySTDinterwordspacing

\bibitem{DT}
\BIBentryALTinterwordspacing
H.~Yang \emph{et~al.}, ``The digital twin framework for the physical wideband and long-haul optical fiber communication systems,'' \emph{Laser Photonics Rev.}, vol.~18, no.~10, p. 2400234, 2024.
\BIBentrySTDinterwordspacing

\bibitem{Transfer_learning}
S.~J. Pan and Q.~Yang, ``A survey on transfer learning,'' \emph{IEEE Trans. Knowl. Data Eng.}, vol.~22, no.~10, pp. 1345--1359, 2010.

\bibitem{many_to_many}
S.~Deligiannidis, C.~Mesaritakis, and A.~Bogris, ``Performance and complexity analysis of bi-directional recurrent neural network models versus volterra nonlinear equalizers in digital coherent systems,'' \emph{J. Lightw. Technol.}, vol.~39, no.~18, pp. 5791--5798, 2021.

\bibitem{Transformer}
\BIBentryALTinterwordspacing
A.~Vaswani \emph{et~al.}, ``Attention is all you need,'' 2023. [Online]. Available: \url{https://arxiv.org/abs/1706.03762}
\BIBentrySTDinterwordspacing

\bibitem{Scaling_law}
\BIBentryALTinterwordspacing
J.~Kaplan \emph{et~al.}, ``Scaling laws for neural language models,'' 2020. [Online]. Available: \url{https://arxiv.org/abs/2001.08361}
\BIBentrySTDinterwordspacing

\bibitem{LSTM_VS_Transformer}
S.~Karita \emph{et~al.}, ``A comparative study on transformer vs rnn in speech applications,'' in \emph{2019 IEEE Automatic Speech Recognition and Understanding Workshop (ASRU)}, 2019, pp. 449--456.

\bibitem{GPT1}
A.~Radford, K.~Narasimhan, T.~Salimans, I.~Sutskever \emph{et~al.}, ``Improving language understanding by generative pre-training,'' 2018.

\bibitem{GPT2}
A.~Radford, J.~Wu, R.~Child, D.~Luan, D.~Amodei, I.~Sutskever \emph{et~al.}, ``Language models are unsupervised multitask learners,'' \emph{OpenAI blog}, vol.~1, no.~8, p.~9, 2019.

\bibitem{GPT3}
T.~Brown \emph{et~al.}, ``Language models are few-shot learners,'' \emph{Advances in neural information processing systems}, vol.~33, pp. 1877--1901, 2020.

\bibitem{Transformer_survey}
\BIBentryALTinterwordspacing
T.~Lin, Y.~Wang, X.~Liu, and X.~Qiu, ``A survey of transformers,'' 2021. [Online]. Available: \url{https://arxiv.org/abs/2106.04554}
\BIBentrySTDinterwordspacing

\bibitem{vit}
\BIBentryALTinterwordspacing
A.~Dosovitskiy \emph{et~al.}, ``An image is worth 16x16 words: Transformers for image recognition at scale,'' 2021. [Online]. Available: \url{https://arxiv.org/abs/2010.11929}
\BIBentrySTDinterwordspacing

\bibitem{BERT}
\BIBentryALTinterwordspacing
J.~Devlin, M.-W. Chang, K.~Lee, and K.~Toutanova, ``Bert: Pre-training of deep bidirectional transformers for language understanding,'' 2019. [Online]. Available: \url{https://arxiv.org/abs/1810.04805}
\BIBentrySTDinterwordspacing

\end{thebibliography}










\newpage

 
\vspace{11pt}


\vspace{11pt}


\vfill
\end{document}